\documentclass[review]{elsarticle}

\usepackage{amsthm}
\usepackage{subfig}
\usepackage[table,x11names]{xcolor}
\usepackage{amsmath}
\newtheorem{mydef}{Definition}

\usepackage{smartdiagram}
\usepackage{graphicx}
\usepackage{soul}
\usepackage{float}
\usepackage{multirow}
\usepackage{booktabs} 
\usepackage{etoolbox} 
\usepackage[vlined,boxed,commentsnumbered, ruled, linesnumbered]{algorithm2e}
\SetKw{KwBy}{by}
\usepackage{tablefootnote}
\makeatletter
\newcommand{\removelatexerror}{\let\@latex@error\@gobble}
\makeatother

\usepackage{smartdiagram}
\usepackage[margin=3cm]{geometry}

\usepackage{footnote}

\usepackage[margin=3cm]{geometry}

\makeatletter
\def\BState{\State\hskip-\ALG@thistlm}
\makeatother

\usepackage{setspace}

\usepackage{breakcites}
\usepackage{enumerate}
\usepackage{makecell}

\usepackage{float}
\usepackage{listings}

\usepackage{tikz}
\usetikzlibrary{trees}
\usepackage{paralist}
\usepackage{ragged2e}
\usepackage[framemethod=tikz]{mdframed}

\usetikzlibrary{positioning,arrows.meta}

\definecolor{arrowblue}{RGB}{98,145,224}

\usepackage{soul}

\usepackage[colorinlistoftodos]{todonotes}

\usepackage{lineno,hyperref}
\usepackage{cleveref}
\usepackage{setspace}
\journal{Computer Communications}

\begin{document}

\begin{frontmatter}

\title{Privacy Preserving Distributed Machine Learning with Federated Learning}

\author[mymainaddress,mysecondaryaddress]{M.A.P.~Chamikara
	\corref{mycorrespondingauthor}}
\cortext[mycorrespondingauthor]{Corresponding author}
\ead{pathumchamikara.mahawagaarachchige@rmit.edu.au}

\author[mymainaddress]{P.~Bertok}
\author[mymainaddress]{I.~Khalil}
\author[mysecondaryaddress]{D.~Liu}
\author[mysecondaryaddress]{S.~Camtepe}

\address[mymainaddress]{RMIT University, Australia}
\address[mysecondaryaddress]{CSIRO Data61, Australia}

\begin{abstract}
\begin{mdframed}[backgroundcolor=green!50,rightline=false,leftline=false]
\centering 
The published article can be found at \url{https://doi.org/10.1016/j.comcom.2021.02.014}
\end{mdframed}

Edge computing and distributed machine learning have advanced to a level that can revolutionize a particular organization. Distributed devices such as the Internet of Things (IoT) often produce a large amount of data, eventually resulting in big data that can be vital in uncovering hidden patterns, and other insights in numerous fields such as healthcare, banking, and policing.  Data related to areas such as healthcare and banking can contain potentially sensitive data that can become public if they are not appropriately sanitized. Federated learning (FedML) is a recently developed distributed machine learning (DML) approach that tries to preserve privacy by bringing the learning of an ML model to data owners'. However, literature shows different attack methods such as membership inference that exploit the vulnerabilities of ML models as well as the coordinating servers to retrieve private data. Hence, FedML needs additional measures to guarantee data privacy. Furthermore, big data often requires more resources than available in a standard computer. This paper addresses these issues by proposing a distributed perturbation algorithm named as DISTPAB, for privacy preservation of horizontally partitioned data.  DISTPAB alleviates computational bottlenecks by distributing the task of privacy preservation utilizing the asymmetry of resources of a distributed environment, which can have resource-constrained devices as well as high-performance computers.  Experiments show that DISTPAB provides high accuracy, high efficiency, high scalability, and high attack resistance. Further experiments on privacy-preserving FedML show that DISTPAB is an excellent solution to stop privacy leaks in DML while preserving high data utility.
\end{abstract}


\end{frontmatter}

\section{Introduction}
\label{intro}

\begin{figure}[H]
\centering
	\includegraphics[width=1\textwidth, trim=0.5cm 0cm 0.5cm 0cm]{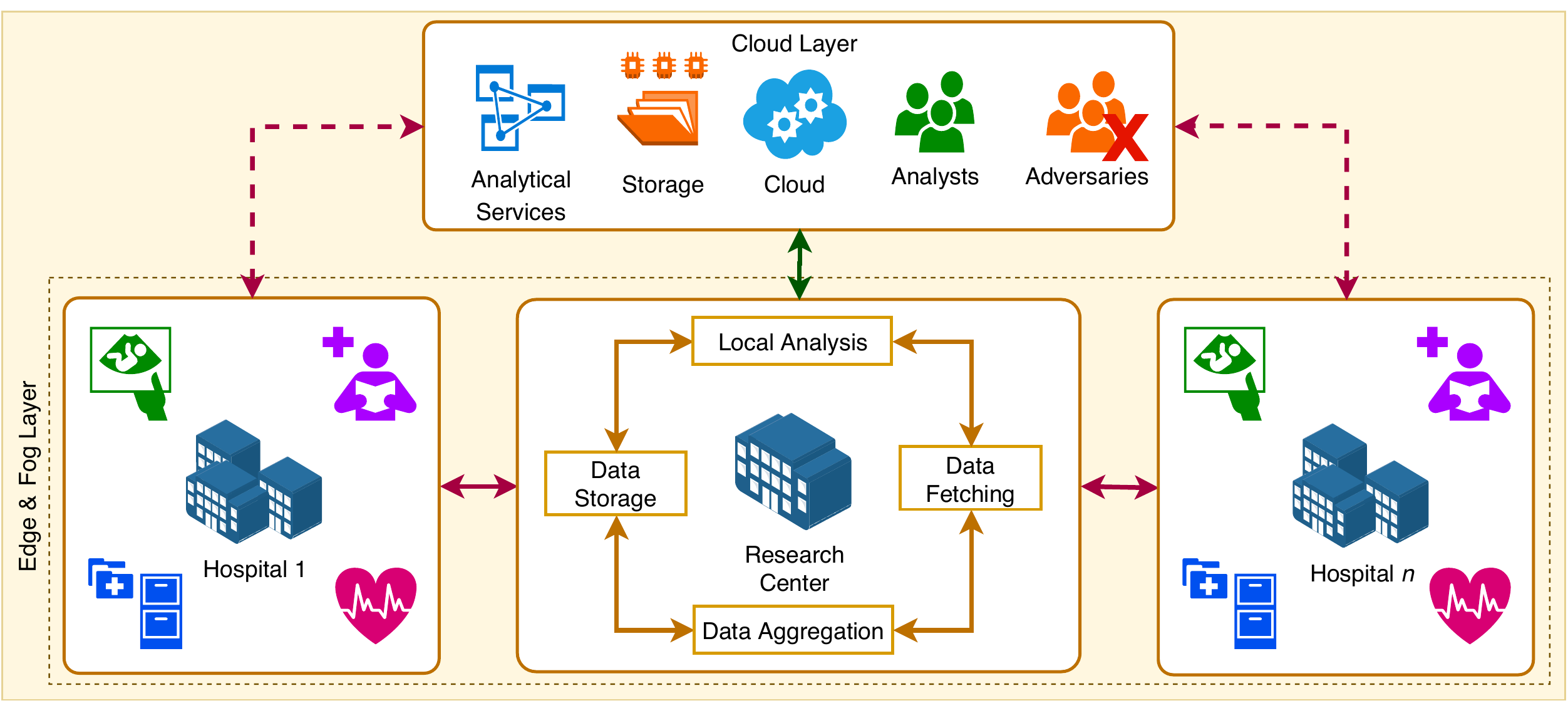}
	\caption{An example for a distributed organizational setting: a healthcare ecosystem which is geographically distributed among multiple locations. A healthcare system may have many distributed branches to it, facilitating and collecting many healthcare data including IoT sensor data. The central body coordinates the distributed hospitals in terms of maintaining data integrity to support a wide range of analytics. The central authority/research centre is also responsible for sharing data to cloud-based third parties for enhanced intelligence and quality of service towards their patients.}
	\label{iotecho}
\end{figure}

The amalgamation of different technologies such as edge computing, IoT, cloud computing, and machine learning has contributed to a rapid proliferation of technological development in many areas such as healthcare and banking~\cite{tegegne2014enriching,kim2017information,serban2019real}. The increase of cheap pervasive sensing devices has contributed to the rapid growth of IoT, becoming one of the main sources of big data~\cite{khan2018iot}. In the broader spectrum of sensor systems, cyber-physical systems and advanced analysis tools are converged together to provide consolidated services. As a result, a particular system  (e.g. healthcare, banking) can now be benefited from multiple sources of data, additionally to what is accumulated by conventional means~\cite{arachchige2020trustworthy}. This growing availability of different sources of data has been able to revolutionize leading fields such as healthcare technologies to achieve excellent achievements in many areas such as drug discovery, early outbreak detection, epidemic control analysis, which were once considered to be complicated~\cite{kim2017information,serban2019real}. However, data related to the fields such as healthcare, banking and policing are massively convoluted with sensitive private data~\cite{arachchige2019local,chamikara2019efficient, chamikara2016fuzzy}. It is essential to go for extreme measures to protect sensitive data while analyzing them to generate meaningful insights~\cite{alabdulatif2018real,alabdulatif2019secure}. However, it is an extremely challenging task as systems related to fields such as healthcare and banking are often densely distributed.  This paper examines the issues related to distributed data sharing and analysis in order to devise an optimal privacy preservation solution towards distributed machine learning~\cite{bonawitz2019towards} in environments such as presented in Figure \ref{iotecho}, which represents a typical distributed industry setup (e.g. smart healthcare, open banking) that runs on IoT, edge, fog, and cloud computing.

Privacy violations in fields such as healthcare and banking can be catastrophic due to the availability of highly sensitive person-specific data~\cite{chamikara2019efficient}.  Among different definitions, privacy for data sharing and analysis can be defined as ``Controlled Information Release'' \cite{bertino2008survey}. It has been shown before that it is easy to identify patients in a database by combining several quasi-identifiers such as age, postcode, and sex~\cite{samarati2001protecting}. Removing just the identifiers from the dataset before releasing is not enough to protect the individuals' privacy, and leaking personal information to untrusted third parties can be catastrophic~\cite{chamikara2018efficient, lopez2013privacy, bilge2013scalable, li2020voluntary}. Privacy-preserving data mining (PPDM) is the area that applies privacy-preserving approaches to data mining methods to protect the private information of the users of the underlying input data during the data mining processes~\cite{chamikara2018efficient}. In this paper, we investigate the PPDM solutions that can be applied to limit privacy leaks in distributed machine learning (DML) under big data settings.  The area of homomorphic encryption is widely explored for PPDM. However, in terms of big data and DML, homomorphic encryption cannot address the three challenges, (i) efficiency, (ii) high volume, and (iii) massive distribution of data. Furthermore, homomorphic encryption increases the data size during the encryption ( e.g. single bit can be multiplied to 16 bits), which is unreliable for big data and increases the data storage burdens ~\cite{zhou2017security}.  Compared to encryption, data perturbation (data modification) can provide efficient solutions towards privacy preservation with a predetermined error that can result due to the data modification~\cite{chamikara2018efficient,yargic2019privacy}.

Federated Learning (FedML) is a distributed machine approach that is developed to provide efficient privacy-preserving machine learning in a distributed environment~\cite{yang2019federated,thapa2020splitfed}. In FedML, the machine learning model generation is done at the data owners' computers, and a coordinating server (e.g. a cloud server) is used to generate a global model and share the ML knowledge among the distributed entities (e.g. edge devices). Since the original data never leave the data owners' devices, FedML is assumed to provide privacy to the raw data.  However, ML models show vulnerability to privacy inference attacks such as model memorizing attacks and membership inference, which focus on retrieving sensitive data from trained ML models even under black-box settings~\cite{song2017machine,shokri2017membership}. Model inversion attacks that recover images from a facial recognition system~\cite{fredrikson2015model} is another example that shows the vulnerability of ML to advanced adversarial attacks. If adversaries gain access to the central server/coordinating server, they can deploy attacks such as model memorizing attacks, which can memorize and extract raw data from the trained models~\cite{song2017machine}. Hence, DML approaches, such as FedML, need additional measures to guarantee that there will not be any unanticipated privacy leaks. Differential privacy is a privacy definition (privacy model) that offers a strong notion of privacy compared to previous models~\cite{arachchige2019local}. Due to the application of heavy noise, the previous attempt to enforce differential privacy on big data has resulted in low utility in terms of advanced analytics, which can be catastrophic for applications such as healthcare~\cite{akgun2015privacy}. A major disadvantage of other techniques such as random rotation and geometric perturbation is their incapability to process high dimensional data (big data) efficiently. These perturbation approaches spend an excessive amount of time to generate better results with good utility and privacy~\cite{chen2005random,chen2011geometric}. In terms of efficiency, additive perturbation provides good performance. However, the perturbed data end up with a low privacy guarantee ~\cite{okkalioglu2015survey}.  Another issue that is often ignored when developing privacy-preserving mechanisms for big data is data capacity issues. The application of privacy preservation (the specific algorithms based on encryption on perturbation) on a large database can be extensive and impossible if the exact resource allocation scenarios are not implemented correctly. A recently developed algorithm named PABIDOT for big data privacy preservation promises to provide high efficiency towards big data~\cite{chamikara2020efficient}. The proposed method shows high classification accuracy for big data while providing high privacy. However, PABIDOT  cannot be used for DML as it does not address the data distribution and data perturbation using resource-constrained devices (as depicted in Figure \ref{iotecho}).

We propose a new DISTributed Privacy-preserving Approach for distriButed machine learning in geographically distributed data and systems~(DISTPAB). DISTPAB is a distributed privacy-preserving algorithm that employs a data perturbation mechanism.   The distributed scenario of data perturbation of DISTPAB allows the perturbation of extensive datasets that need to be shared among distributed entities without leaking privacy.  The actual data perturbation is conducted in the distributed entities (in edge/fog devices) locally using the global perturbation parameters generated in a central coordinating node before conducting FedML. This way, DISTPAB restricts original data to be communicated  (before perturbation) via the network, which can be attacked by adversaries. The global perturbation parameter generation of DISTPAB ensures that there is no degradation of accuracy or attack resistance of the perturbed data. DISTPAB was first tested using six datasets obtained from the data repository named "UCI Machine Learning Repository"\footnote{http://archive.ics.uci.edu/ml/index.php}. The results show that DISTPAB provides excellent classification accuracy, attack resistance, and excellent efficiency towards distributed machine learning under big data settings.

The following sections of the paper are set out as follows.  Section \ref{litrev} includes a summary of related work. Section \ref{fndmntls} provides a summary of the fundamentals used in developing DISTPAB. Section \ref{ourapproach} provides background information related to DISTPAB. Section \ref{resdiscussion} presents the experimental evaluations of the the performance and the attack resistance of DISTPAB. Section \ref{discussion} provides a discussion on the results provided in Section \ref{resdiscussion}. The paper is concluded in Section \ref{conclusion}.

\section{Literature Review}
\label{litrev}

Distributed systems such as available in healthcare have become vastly complex due to the amalgamation of different technologies such as IoT,  edge computing, and cloud computing. Due to these advanced capabilities, a modern system can utilize a myriad of data sources to facilitate improved capabilities towards essential services. The nature of distributed data platforms introduces a plethora of complexities towards preserving user privacy of users without compromising data utility. Extremely high dimensions and massive distribution of data sources are two of the main complexities that need to be addressed when designing robust privacy-preserving approaches for DML systems.  Due to these reasons, privacy-preserving data mining (PPDM) approaches should be efficient and should be able to work in distributed settings. Approaches based on secure multi-party computation~\cite{oleshchuk2009internet}, attribute access control ~\cite{khan2018iot}, lightweight cryptographic procedures ~\cite{zhou2017security}, homomorphic encryption~\cite{zhou2017security} are few examples for some encryption approach which can provide good privacy. However, the high computational complexity of such encryption scenarios can seriously jeopardize the performance of a distributed ML setup. The inherent high computational complexity of the cryptographic approaches requires excessive amounts of computational resources, which include large storage facilities such as cloud computing that are often controlled by third-parties. Furthermore, the encryption approaches such as homomorphic encryption for PPDM increase the size of the data after the application of encryption, which is not suitable for the domain of big data as the data sizes can already be extensive. Hence, the implementation of cryptographic scenarios for distributed databases can be unrealistic and unaffordable~\cite{akgun2015privacy}.   Compared to encryption, data perturbation/modification provides efficient and lightweight solutions for big data privacy~\cite{okkalioglu2015survey}, and hence, data perturbation is a better fitting solution for  PPDM of distributed data.

Previous data perturbation approaches include swapping ~\cite{hasan2016effective}, additive perturbation ~\cite{muralidhar1999general},  condensation~\cite{aggarwal2004condensation}, randomized response~\cite{fox2015randomized},  microaggregation~\cite{soria2015t},  random rotation ~\cite{chen2005random}, random projection ~\cite{liu2006random}, geometric perturbation~\cite{chen2011geometric}, and hybrid perturbation ~\cite{aldeen2015comprehensive}.   Due to the modifications, the data end up with reduced utility. The relationship between utility and privacy granted by a particular perturbation algorithm is defined via a privacy model~\cite{machanavajjhala2015designing}. More than a few privacy models have been introduced where one model tries to overcome the defects of another. $k-anonymity$~\cite{niu2014achieving, navarro2012user},  $l-diversity$~\cite{machanavajjhala2006diversity},  $t-closeness$~\cite{li2007t},  $(\alpha, k)-anonymity$~\cite{wong2006alpha}, $k_{\theta}-affinity$~\cite{carpineto2015ktheta}  are some examples for privacy models. However,  these models exhibit vulnerabilities to attacks such as composition attacks~\cite{ganta2008composition}, minimality attacks ~\cite{zhang2007information},  and foreground knowledge~\cite{wong2011can} attacks. Moreover, the perturbation approaches that use these models do not scale for big data with high dimensions. When the dimensions of the input dataset grow, the computational cost necessary to conduct data perturbation grows exponentially. The literature refers to this phenomenon as "The Dimensionality Curse" ~\cite{aggarwal2008privacy}.  Another issue in high dimensional data is the leak of extra information that can be effectively used by attackers to misuse sensitive private information~\cite{bettini2015privacy}. Differential privacy (DP) is another privacy model, which was developed more recently to render a strong privacy guarantee compared to prior approaches.  Nevertheless, due to the high noise levels imposed by DP algorithms, DP might not be the optimal choice for big data privacy~\cite{akgun2015privacy}.

The distribution of data sources and infrastructures also introduces a massive complexity towards the development of robust privacy-preserving approaches for distributed databases. Modern distributed privacy-preserving approaches include federated (FedNN) machine learning and split learning (SplitNN), which can provide a certain level of privacy to distributed databases.  In FedML, the distributed clients use their local data to train local ML models and communicate the locally trained ML model parameters with the central server to train a global representation of the local models. The server then distributes the global model parameters with the clients so that the clients can now generalize their local models based on the model parameters federated by the server. SplitML has a similar distributed setup. However, instead of communicating the model parameters, SplitML splits the ML model between the clients and the server. During the training process, a client will hold a portion of the ML model, whereas the server will hold the other portion of the ML model. As a result, a client will transfer activations from its split layer to the split layer of the server during both forward and backward passes of the training process of the ML model. When many clients need to connect to the server for ML model training, the clients will either have a peer to peer communication or a client-server communication for the model parameter communications in order for each client to have a synchronous ML model training process. However, both FedML and SplitML have the same problem of the central point of failure as the server in both cases has too much control over the model learning process. If the server is attacked, the whole framework becomes vulnerable, and as a result, user privacy cannot be guaranteed. Especially the attack methods such as membership inference and model memorization can exploit the vulnerabilities of the central point of failure to retrieve raw data used for the model training process. The other distributed approaches which work on horizontally and vertically partitioned data also tend to produce issues towards efficiency or privacy~\cite{hardy2017private}. As most of these methods use the approaches based on homomorphic encryption to provide privacy, such methods become inefficient when working with big data, and the need to share too much information with the server makes most of them untrusted and yields the same issue of the central point of failure~\cite{hardy2017private}.

Among data perturbation approaches, matrix multiplicative approaches proved to provide high utility towards data clustering and classification~\cite{chamikara2018efficient,chamikara2020efficient}.  Examples for matrix multiplicative perturbation approaches include random rotation, geometric, and projection perturbation~\cite{okkalioglu2015survey} methods.  For example, random rotation perturbation repeatedly multiplies the input data by a randomly generated matrix with the properties of an orthogonal matrix, until the perturbed data satisfy a pre-anticipated level of privacy ~\cite{chen2005random}. An additional matrix of translation and distance perturbation are combined with random rotation to produce geometric data perturbation. The added randomness of the random translation matrix improves the privacy of the perturbed data. However, geometric data perturbation also follows the same repeated approach until the perturbed data satisfy an expected level of privacy~\cite{chen2011geometric}. Random projection perturbation follows a different approach by projecting the high dimensional input data to a low dimensional space ~\cite{liu2006random}. One of the important properties of matrix multiplicative data perturbation approaches is their ability to preserve the distance between tuples of the input dataset~\cite{chen2005random,chen2011geometric,liu2006random}. As a result of this property, matrix multiplicative approaches provide high utility for data classification and clustering, which are based on distance calculations. However, due to the inefficient approaches utilized for the optimal perturbation parameter generation, rotation perturbation, geometric perturbation, and projection fail to provide enough scalability. PABIDOT is a recently developed perturbation approach for the privacy preservation of big data classification, using the properties of geometric data perturbations. PABIDOT provides high efficiency towards big data while maintaining high privacy and classification accuracy. However, for distributed healthcare scenarios such as shown in Figure \ref{iotecho}, PABIDOT cannot be applied as it is not a distributed algorithm.  As shown in Figure \ref{iotecho}, a distributed system can be composed of branches that are geographically dispersed. The data coming from each branch should be perturbed before they leave the local network. A distributed data perturbation algorithm which can provide high utility while maintaining high privacy for distributed healthcare is essential. 

\section{Fundamentals}
\label{fndmntls}
The proposed method uses multidimensional transformations and $Randomized$ $Expansion$~\cite{chamikara2020efficient}, which improves the randomness of the data. DISTPAB considers the input dataset as a data matrix in which each tuple is regarded as a column vector for applying the transformations. This section explains how the data transformation is for perturbation and how the randomization of the final output is improved using $Randomized$ $Expansion$.

\subsection{Data matrix (D)} 
We consider the input dataset as a data matrix. The proposed perturbation mechanism only addresses numeric data perturbation. Hence a data matrix will contain only numeric data. The columns of the data matrix represent the attributes/features, whereas the rows (row vectors) represent the member (data owner) records. Each row vector will be subjected to multidimensional transformations on an $n$-dimensional Cartesian coordinate ($n$ = number of attributes)  system.

\subsection{Multidimensional isometric transformations}
If a multidimensional transformation, $\ T: R^n\rightarrow R^n $ holds Equation \ref{isometrictrans}. Reflection, translation and rotation are examples for multidimensional isometric transformation~\cite{maruskin2012essential}.

\begin{equation}
|T(A)-T(B)|=|A-B|,\quad\forall~A,B\in R^n
\label{isometrictrans}
\end{equation}

\subsection{Homogeneous coordinate form and composite operations}
In ${n-dimensional}$ space, we can write a homogeneous coordinate point as an $\ (n+1) $ dimensional position vector $\ (x_1,x_2,\dots,x_n,h)$, where $\ h \neq 0$ is an additional term.

An $\ (n+1) $ dimensional position vector $\ (x_1,x_2,\dots,x_n,h)$, represents a  homogeneous coordinate point in the  ${n-dimensional}$ space when $\ h \neq 0$ is an additional term. This coordinate form allows representing all transformations in matrix multiplication form of size of $\ (n + 1)\times(n + 1) $~\cite{jones2012computer}.  

A transformation operation is called a composite transformation when it involves the operations between several matrices. Equation \ref{comptrans} shows the composite transformation of the sequential application of $\ M_1, M_2, M_3,\dots $ to a homogeneous matrix $\ X $.

\begin{equation}
X'=\dots(M_3(M_2(M_{1}X)))
=\dots M_3\times M_2\times M_1\times X
\label{comptrans}
\end{equation}
  
We can use a new column of ones, where $h=1$, to convert a data matrix into its homogeneous matrix form~\cite{chamikara2020efficient}.

\subsection{z-score normalization of the input dataset, $D$}
The attributes of an input dataset can have different scales, which requires different levels of perturbation to each attribute. Consequently, applying the same levels of perturbation would not provide equal protection to all the attributes. As a result, the final version of the perturbed dataset will have attributes with insufficient levels of perturbation. We apply z-score normalization (also referred to as standardization) to impose equal weights to all the attributes during the transformations. The z-score normalization scales the input dataset with a standard deviation equals to $1$ and a mean equals to $0$~\cite{kabir2015novel}.

\subsection{Generating transformation matrices for perturbation}
For the perturbation of the input data matrix, we need to generate the n-dimensional homogeneous translation, reflection, and rotation matrices. We can generate the translational matrix according to  ~\cite{jones2012computer} (refer \ref{apptrans}). Since the input data matrix is z-score normalized, the translational coefficients of the translational matrix are sampled from a uniform random distribution that lies in the interval $(0,1)$~\cite{chamikara2020efficient}. We can generate the (n+1) axis reflection matrix utilizing  n-dimensional reflection matrix (refer to \ref{appreflect}) ~\cite{chamikara2020efficient}. The rotational matrix can be generated using the concept of concatenated subplane rotation (refer \ref{approtate})~\cite{chamikara2020efficient}. 

\subsection{Randomized expansion}
\label{ranexp}
Randomized expansion is a noise addition approach which improves the randomness of the perturbed data~\cite{chamikara2020efficient}.  Equation \ref{randexp} shows the randomized expansion noise generation where $S_{\pm}$ is the sign matrix generated based on the values of $D^{p2}$ (intermediate perturbed data matrix). As the equation shows, the positiveness or the negativeness is improved by randomized expansion to provide high utility with improved randomization~\cite{chamikara2020efficient}. 

\begin{equation}
\begin{aligned}
D^{p2}=(\left\| D^{p2}\right\|+\left\| \mathcal{N}(0,\sigma)\right\|)\bullet S_{\pm}
\label{randexp}
\end{aligned}
\end{equation}

\subsection{Quantification of privacy}
$Var(D-D^p)$  is one approach to measure the level of privacy of the perturbed data, where $D$ is original data and $D^p$ is perturbed data~\cite{muralidhar1999general}.   The higher the $Var(P)$, the higher the difficulty, hence the higher the privacy. Take $\ X^p$ to be a perturbed data series of attribute $X $. Now, $Var(P)$ can be written as in Equation \ref{varp}, where $P=(X^p-X)$. 

\begin{equation}
Var(P)=Var(p_1,p_2,\dots,p_n)={\frac{1}{n}}\displaystyle\sum_{i=1}^{n}(p_i-\bar{p})^2
\label{varp}
\end{equation}

\subsection{$\Phi-separation$}

$\Phi-separation$ is a privacy model that allows the selection of optimal perturbation parameters for a perturbation algorithm in a given instance of data. $\Phi-separation$  allows the determination of the best perturbation parameters while providing an optimal empirical privacy guarantee~\cite{chamikara2020efficient}. 

\begin{mydef}[$\Phi-separation$]

Apply a perturbation algorithm $M$ to the dataset $D=[d_1,d_2,\dots, d_n]_{m\times n}$ to generate the perturbed instances of $D$ as $D_j^p=[d^p_{j1},d^p_{j2},\dots, d^p_{jn}]_{m\times n}$ for $j=1,2,\dots, k$. If $k$ represents the number of all feasible ways to apply $M$ to $D$ to generation $D^p$. Then,

\begin{equation}
\label{phieq1}
\phi_j=min\{[var(d_i-d^p_{ji})]\} \forall j={1,2,\dots,k}
\end{equation}

where $\phi_j$  is the minimum privacy guarantee.

\noindent We can maximize all the possible minimum privacy guarantees ($\phi_j$) to obtain the optimal privacy guarantee $\Phi$ as

\begin{equation}
\label{phieq2}
\Phi=max\{\phi_j\}_{j=1}^k
\end{equation}
A perturbed dataset provides $\Phi-separation$ if it satisfies Equation \ref{phieq2}.
\end{mydef}

\subsection{Federated Learning}
\label{fedlearning}

Federated learning provides the capability to train a machine learning model using distributed data without sharing the original data between the participating entities. Assume that $\left\{D_{1}, \ldots D_{N}\right\}$ are distributed datasets, which are distributed among $N$ data owners $\left\{O_{1}, \ldots O_{N}\right\}$. In a federated learning setup, each of these datasets ($D_{i}$)  never have to leave the corresponding owner ($O_{i}$) and each owner trains ML models locally without exposing the corresponding data to any external parties. The model parameters of the locally trained models are then collected in a server (a central entity/authority), which federate the parameters to generate a global representation of all the models $(ML_{fed})$. The accuracy $(A_{fed})$ of $ML_{fed}$ should be very close to the accuracy $(A_{ctr})$ of the model trained centrally with all the data  ~\cite{yang2019federated}. This relationship can be represented using Eq. \ref{fedctr}, where $\delta$ is a non-negative real number.

 \begin{equation}
\left|A_{fed}-A_{ctr}\right|<\delta
\label{fedctr}
\end{equation}

\subsubsection{Horizontal and vertical federated learning}

In horizontal federated learning, all the clients share the same feature space but different space in samples. Consider a feature space $X: x_1,\dots,x_n$, a sample ID space of $I: i_1, \dots,i_j$, and a label space of $Y: y_1,\dots,y_k$ constitutes the complete training dataset $(I,X,Y)$. Then a particular distributed client will have the dataset $(i_l,X,Y)$. Several regional banks, with different user groups with similar business characteristics having the same feature spaces, can be an example of a horizontally partitioned scenario. In vertical federated learning, all the clients share the same sample ID ($I$) space and different feature spaces ($x_1,\dots,x_n$). Then a particular distributed client will have the dataset  $(I,x_l,Y)$. Different companies with different businesses having different feature spaces and the same user group, is an example of a vertically partitioned scenario. In this work, we consider only the horizontal federated learning scenario. 

In the next section, we provide a detailed description of how these fundamentals are used in developing the proposed approach (DISTPAB) for distributed machine learning. 

\section{Our Approach: DISTPAB}
\label{ourapproach}

\begin{figure}[H]
\centering
	\includegraphics[width=0.5\textwidth, trim=0.5cm 0cm 0.5cm 0cm]{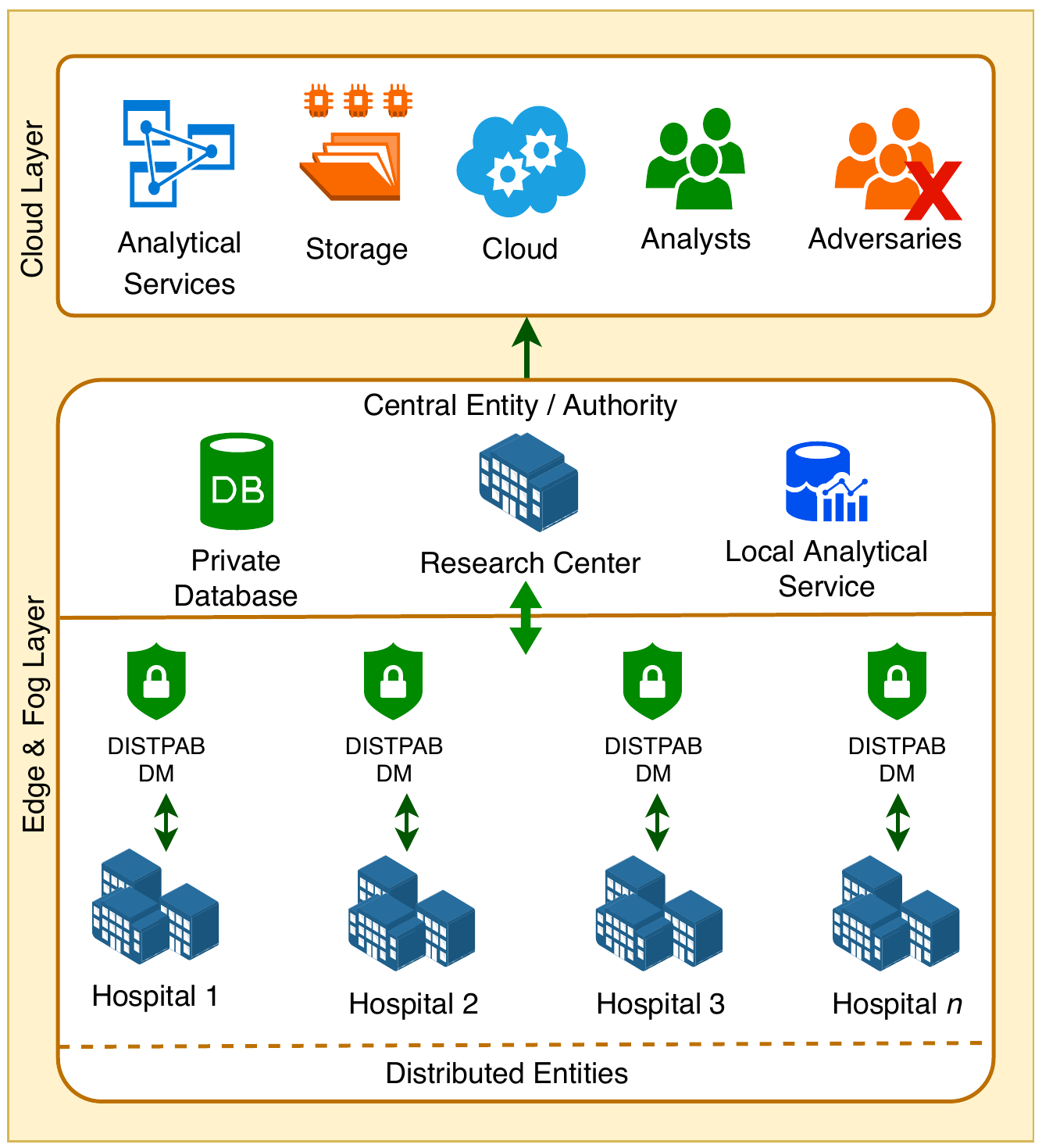}
	\caption{An example scenario where DISTPAB is integrated to a distributed healthcare system to preserve the privacy of data.  The lowest layer represents the distributed set of entities (hospitals) and their interaction with DISTPAB within the fog and edge bounds. As the figure shows, the distributed components of DISTPAB are integrated into each of the distributed hospitals for performing the data perturbations.  The central controlling entity (research centre) will be communicating with the cloud layer, which supports third-party analytical services. In this setup, we assume that no party saves original IoT data, and will only save the perturbed data.}
	\label{iotechodist}
\end{figure}

This section explains the steps of developing a distributed data perturbation algorithm (named as DISTPAB)  for a distributed data and machine learning.  Figure \ref{iotechodist} shows the application of privacy to a distributed healthcare ecosystem shown in Figure \ref{iotecho}. As shown in the figure, the main goal of DISTPAB is to shift the perturbation to the distributed branches before the data leave the local edge and fog layer. However, in doing so, the algorithm should not lose global utility. To achieve that, the main branch (the research center) and the distributed branches will conduct a perturbation parameter exchange. As the perturbation is done using the globally optimal parameters, DISTPAB can maintain a proper balance between utility and privacy. During our explanations, we will be using the words node and entity alternatively, representing the same concept. However, more specifically, we consider an entity to be a fully populated institute, whereas a node to be one or more computing/processing devices within an entity.

The proposed algorithm delegates data maintenance and perturbation to the distributed entities, leaving only the global perturbation parameter generation to the central entity. Figure \ref{distributed} shows the distributed architecture of DISTPAB, which can be used for the perturbation of healthcare data. In this way, the distributed branches do not have to share the original data with any untrusted third party. However, due to the coordination of the proposed algorithm in generating the global perturbation parameters, the perturbed data can provide high utility.

\subsection{Centralized perturbation paradigm} 

DISTPAB first applies geometric transformations in the order of (1) reflection, (2) translation, and (3) rotation to an input dataset. Next, DISTPAB performs randomized expansion and random tuple shuffling to enhance the randomness of the perturbed data~\cite{chamikara2020efficient}. The perturbation is repeated until the perturbed dataset satisfies $\Phi-separation$. In order to devise the distributed paradigm in DISTPAB, we discuss the centralized approach, which is given in Algorithm \ref{parallelalgo}. Algorithm \ref{parallelalgo} takes only two inputs: (1) the original dataset, (2) the standard deviation of the normal random noise generated for randomized expansion.

Before applying the composite geometric transformations, the algorithm derives the best perturbation parameters using the z-score normalized input data matrix.  However, obtaining the best perturbation parameters at $\Phi$ for a big dataset can be extremely inefficient, as it involves running $D^{p1}=(M_i\times TN^{noise}\times RF_{ax}\times (D^N)^T)^T$ and $\phi_{i,ax}=min(Var(x_{ij}-p_{ij})_{i=1}^m)_{j=1}^n$  $\forall$ $x_{ij}\in D^N$ for  multiple perturbation instances (under the loops in Step \ref{al3step7} and \ref{al3steploop2} of Algorithm \ref{parallelalgo} ) of the original datasets. We can prove that these two steps are equal to $ \phi_{i,ax}=min(\vec1+diag(M_i\times RF_{ax}\times CD^N\times RF_{ax}\times M_i^T)+sum((CD^N\times RF_{ax} \bullet M_i)^T)^T)$ ~\cite{chamikara2020efficient} which is much simpler in time complexity as it uses the corresponding covariance matrix to determine the best perturbation parameters instead of browsing through the whole dataset in each iteration.

The algorithm maximizes the $\phi$ value in each iteration until it obtains $\Phi$ as given in Equation \ref{fthetamax}. The number of perturbed instances of $D$ can be given as, $\ D_{1,1}^p, D_{1,2}^p,\dots, D_{1,179}^p,D_{2,1}^p, D_{2,2}^p,\\ \dots, D_{2,179}^p,\dots,D_{n,1}^p, D_{n,2}^p,\dots, D_{n,179}^p$ where $n$ represents the number of attributes, consequently, the axis of reflection varies from 1 to n. $0<\theta<\pi$, where $\quad\theta \notin \left\{ \pi/6, \pi/4, \pi/3, \pi/2, 2\pi/3, 3\pi/4, 5\pi/6  \right\} $. This will result in a matrix of $\phi$ values (local minimum privacy guarantees) as represented in Equation \ref{phivals}.

\begin{equation}
\begin{bmatrix}
\phi_{1,1} & \phi_{1,2} & \phi_{1,3} & \dots & \phi_{1,179} \\
\phi_{2,1}  & \phi_{2,2} & \phi_{2,3}  & \dots & \phi_{2,179} \\
\vdots  & \vdots & \vdots & \ddots & \vdots  \\ 
\phi_{n,1} & \phi_{n,2} & \phi_{n,3} & \dots & \phi_{n,179} \
\end{bmatrix}_{n\times 179}
\label{phivals}
\end{equation}
As given in Equation \ref{globephi}, we obtain the minimum value from each column of Equation \ref{phivals} to identify the best  $\phi_j$ for each instance of $\theta$.

\begin{equation}
\begin{bmatrix}
\phi_{1} & \phi_{2} & \phi_{3} & \dots & \phi_{179} \\
\end{bmatrix}
\label{globephi}
\end{equation}

Now we choose $\Phi$, as the largest of global minimum privacy guarantees ($\phi_j$) as represented in  Equation \ref{fthetamax}. This can also be explained as finding the highest privacy guarantee $\Phi $ that can be rendered by the most vulnerable attribute in the database.

\begin{equation}
\Phi=max([[\phi_{j}]_{j=1}^{179}])
\label{fthetamax}
\end{equation}

After determining the best perturbation parameters, the geometric transformations are conducted on the normalized input dataset following the order of reflection, noise translation, and rotation. We follow this order of transformations to avoid the points (in the n-dimensional space) closer to the origin, getting lower levels of perturbation. Rotation has a larger effect on the points that are far away from the origin compared to those which are close to the origin. As a result, the points which are close to the origin can get easily attacked. Equation \ref{transformationinstance} shows the order of the application of the three transformations on the normalized input data matrix, where rotation is conducted as the last transformation for the highest perturbation possible. As the next step, we add noise using randomized expansion.  Reverse z-score normalization on the perturbed data is used to generate an output in the attribute ranges similar to the original dataset. Next, the tuples are randomly swapped in order to limit vulnerability to data linkage attacks.

\begin{equation}
D^{p}=(M\times T_{ND}\times RF_{\overline{ND}}\times (D^N)^T)^T
\label{transformationinstance}
\end{equation}

\begin{center}
    \scalebox{0.8}{
    \begin{minipage}{1.1\linewidth}
     \removelatexerror
      \begin{algorithm}[H]
	\caption{Centralized perturbation algorithm}\label{parallelalgo}
			 \KwIn{
			\begin{tabular}{l c l} 
				$D$  & $\gets $ & original dataset\\
                $\sigma$ & $\gets $ & input noise standard deviation \\& & (default value=0.3)\\
			\end{tabular}
			}
			\KwOut{
			\begin{tabular}{ l c l } 
				$D^{p}$  & $\ \gets $ & perturbed dataset  \\
			\end{tabular}
			}
			 $\Phi=0$\;
			 $\theta_{optimal} = 0$ \;
			 $Rif_{optimal}=0$ \;
			 generate $D^{N} $\;
             generate $Cov(D^{N})$\;
			 generate $TN^{noise} $\;
			\For{\textbf{each} $ax$ in  $\left\{1,2,\dots,n\right\}$}{ \label{al3step7}
			 generate $RF_{ax} $ \;
			\For{\textbf{each} $\theta_i $}{ \label{al3steploop2}
			 generate $M_i $ using Algorithm \ref{rotmatgenerate}\;
             $\ \phi_{i,{ax}}=min(\vec 1+diag(M_i\times RF_{ax}\times\; Cov(D^{N})\times RF_{ax}\times M_i^T)+sum((Cov(D^{N})\times RF_{ax} \bullet M_i)^T)^T)$ \label{al2step11}	
			}			
			} \label{al3step18}
			\For{\textbf{each} $\theta_i $}{ 
               $\ \phi_{i}=min(\phi_{i,{ax}})$  where, $ax$ $\in$  $\left\{1,2,\dots,n\right\}$\;	
			}			
             $\Phi = max(\phi_{i})$ where, $i$ $\in$  $\left\{1,2,\dots,\theta \right\}$\;
             $\theta_{optimal}=\theta_i$  at $\Phi $\;
			 $Rif_{optimal}=ax$  at $\Phi $\;
            
			 generate $M\theta $ \;
			 generate $RF_{optimal} $ \;
			$D^{pt}=(M\theta \times TN^{noise}\times RF_{optimal}\times (D^N)^T)^T$ \label{al3step21}\;
            $D^{pt}=(\left\| D^{pt}\right\|+\left\| \mathcal{N}(0,\sigma)\right\|$)$\bullet S_{\pm}$ \label{al2step22}\;
			 $D^{p} $=$\ D^{pt} \bullet STDVEC$+$MEANVEC$\;
			 randomly swap the tuples of $\ D^{p} $ \label{al3swapstep}\; \label{al2step24}
	\end{algorithm}
    \end{minipage}%
    }
  \end{center}

\subsection{Distributing the perturbation}
\begin{figure}[H]
\centering
	\includegraphics[width=0.8\textwidth, trim=0.5cm 0cm 0.5cm 0cm]{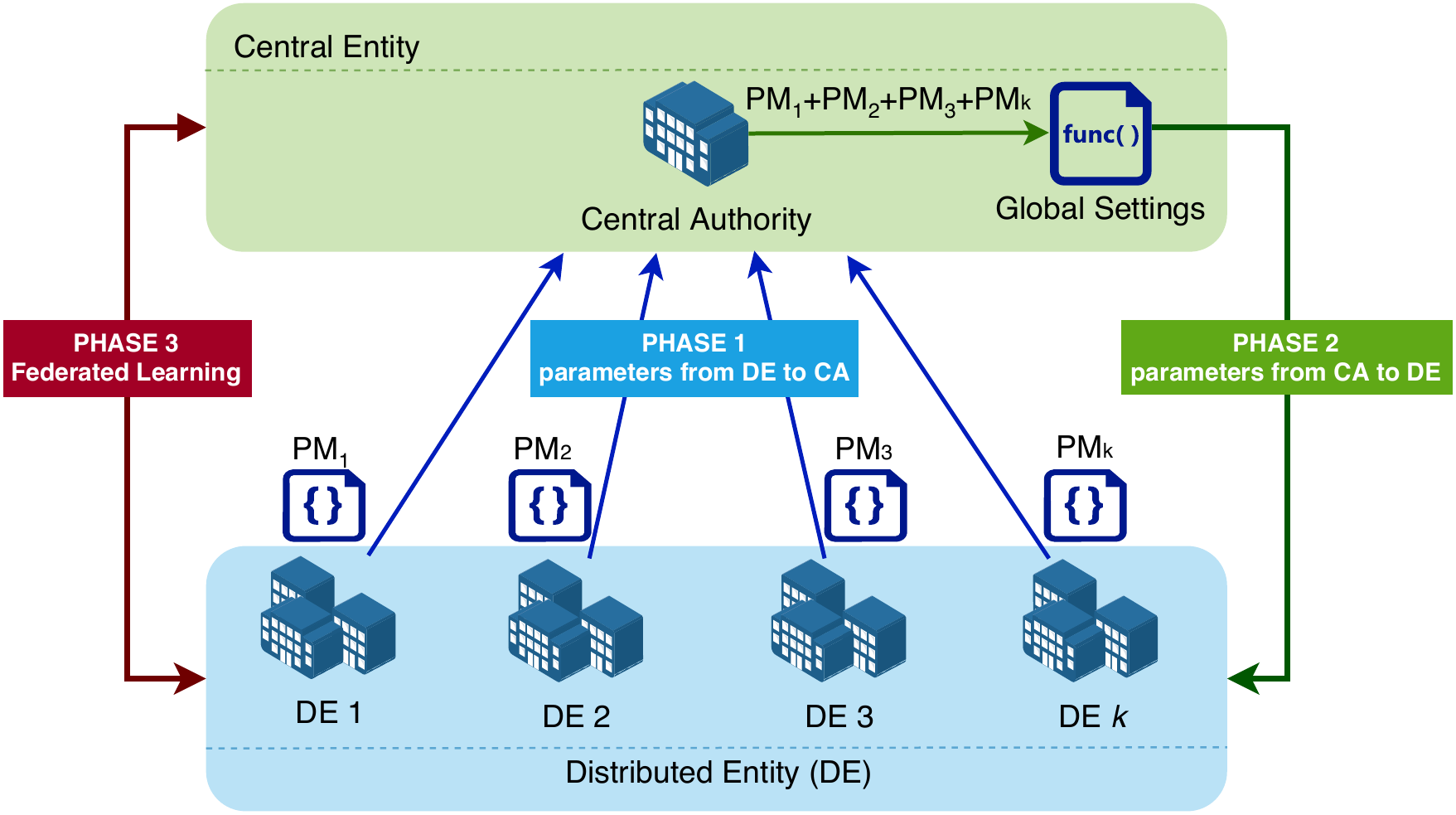}
	
	\caption{Distributed architecture of the perturbation algorithm. The figure shows the abstract view of the distributed perturbation scenario over distributed machine learning. There can be $k$ number of distributed branches that are communicating with the central entity (coordinating server). There are two phases of parameter communication. Phase 1 involves the distributed branches sending local parameters sending to the central entity, whereas Phase 2 involves the central node sending the optimal perturbation parameters (which are calculated based on the local parameters which were sent by the distributed branches) to the distributed branches for perturbation.}
	\label{distributed}
\end{figure}

In order to distribute the perturbation among distributed entities, we break the steps of Algorithm \ref{parallelalgo} into two main phases, as shown in Figure \ref{distributed}.  (1) Generate the global perturbation parameters using the local properties forwarded to the central entity (research centre). (2) Conduct perturbation of data at the distributed entities using the global parameters. (3) Conduct machine learning on perturbed data. In order to achieve these three goals, first, the variance-covariance matrix of each partition of the dataset is passed to the central node, which will run Algorithm \ref{distributedmethod}.  Assume that there are $k$ distributed branches which have the data partitions $\ D_1, D_2,\dots, D_k $. Take $D$ to be the dataset created by merging all the $k$ datasets, as given in Equation \ref{merged}.

\begin{equation}
D = merge(D_1, D_2, \dots, D_k)
\label{merged}
\end{equation}

To merge the covariance matrices, the pairwise covariance update formula introduced in \cite{bennett2009numerically} is adapted. The pairwise covariance update formula for the two merged two column ($u$ and $v$) data partitions, $A$ and $B$, can be written as shown in Equation \ref{covupdate} where the merged dataset is denoted as $\ X $.

\begin{equation}
\resizebox{0.8\textwidth}{!}{$\displaystyle
Cov(X)=\frac{\frac{C_A}{(m_A-1)}+\frac{C_B}{(m_B-1)}+(\mu_{u,A}-\mu_{u,B})(\mu_{v,A}-\mu_{v,B}).\frac{m_A.m_B}{m_X}}{(m_X-1)}
$}
\label{covupdate}
\end{equation}

where $\mu_{u,A}, \mu_{u,A}, \mu_{v,A}, \mu_{v,B}$ are means of $u$ and $v$ of the two data partitions $\ A $ and $\ B $ respectively.   $C_A$ and $C_B$ are the co-moments of the two data partitions $A$ and $B$ where the co-moment of a two column ($u$ and $v$) dataset $D$ is represented as

\begin{equation}
C_D=\sum_{(u,v)\in D} (u-\mu_u)(v-\mu_v)
\end{equation}

Therefore, the variance-covariance matrix update formula of the two data partitions $\ D_1 $ and $\ D_2 $ can be written as shown in Equation \ref{matcovupdate}.

\begin{equation}
\resizebox{0.9\textwidth}{!}{$\displaystyle
Cov(D_{new})=\frac{\frac{Cov(D_1)}{(m_{D_1}-1)}+\frac{Cov(D_2)}{(m_{D_2}-1)}+(\mu_{D_1}(MI_1)-\mu_{D_2}(MI_1))(\mu_{D_1}(MI_2)-\mu_{D_2}(MI_2)).\frac{m_{D_1}.m_{D_2}}{m_{D_{new}}}}{(m_{D_{new}}-1)}
$}
\label{matcovupdate}
\end{equation}

where $\ Cov(D_{new}), Cov(D_1), Cov(D_2) $ are the covariance matrices of the merged dataset $\ D_{new} $ and the data partitions $\ D_1 $ and $\ D_2 $ respectively. $\mu_{D_1} $ and $\mu_{D_2} $ are mean vectors of $\ D_1 $ and $\ D_2 $ respectively and 

\begin{equation}
MI_1=
\begin{bmatrix}
[1]_n\\
[2]_n\\
[3]_n\\
\vdots\\
[n]_n\\
\end{bmatrix}_{n\times n}
\end{equation}

\begin{equation}
MI_2 = (MI_1)^T 
\end{equation}

After generating the covariance matrix ($Cov(D)$)  based on the global parameters (refer Equation \ref{globalcov}), we can produce the vector of standard deviations ($STDVEC=\sqrt{diag(Cov(D))}$) based on the diagonal vector of $Cov(D)$. We can generate the vector of means ($MEANVEC$) using Equation \ref{meanvect} where $\Bar{y_i}=Mean(D_i)$. 

\begin{equation}
Cov(D)=
\begin{bmatrix}
Var_{1} & Cov_{1,2} & \dots & Cov_{1,n} \\
Cov_{2,1}  & Var_{2} & \dots & Cov_{2,n} \\
\vdots  & \vdots & \ddots & \vdots  \\ 
Cov_{n,1} & Cov_{n,2} & \dots & Var_{n} \
\end{bmatrix}_{n\times n}
\label{globalcov}
\end{equation}

\begin{equation}
    MEANVEC=\frac{m_1\times\overline{y_1}+m_2\times\overline{y_2}+\hdots+m_k\times\overline{y_k}}{m_1+m_2+\hdots +m_k}
    \label{meanvect}
\end{equation}

Each distributed branch needs to generate the covariance matrix ($Cov(partition_i)$) and the mean vector ($Mean(partition_i)$) of the corresponding data partition and pass it to the central entity for the execution of Algorithm \ref{distributedmethod}. After completing the execution of Algorithm \ref{distributedmethod}, the central node will pass the required parameters to the distributed entities according to Algorithm \ref{nodeworkload}, as shown in Figure \ref{distributed}. Table \ref{datacom} shows the parameter exchange between the central entity and the distributed entities during the perturbation process.

\begin{table}[H]
\caption{Data (parameter values) passed between the central entity and the distributed entities in the two phases of communications. As shown in Figure \ref{distributed}, in phase 1, the distributed entities send the local perturbation parameters to the central entity. In phase2, the central entity sends the global parameters (which are calculated using the local parameters) to the distributed entities.}
\label{datacom}
\centering 
    \resizebox{0.7\columnwidth}{!}{
    \begin{tabular}{l l}
\toprule
\textbf{\begin{tabular}[c]{@{}l@{}}Phase 1: from distributed\\ nodes to central node\\ \\ (Local parameters related to\\ each data partition)\end{tabular}} & \textbf{\begin{tabular}[c]{@{}l@{}}Phase 2: from centralized node\\ to distributed nodes\\ \\ Global parameters related to\\ all the partitions\end{tabular}} \\ \midrule
1. covariance matrix                                                                                                                                            & 1. random reflection matrix                                                                                                                                   \\ 
2. vector of means                                                                                                                                                 & 2. random translation matrix                                                                                                                                  \\ 
\multirow{2}{*}{3. Number of attributes}                                                                                                                        & 3. random rotation matrix                                                                                                                                     \\ 
                                                                                                                                                                & \begin{tabular}[c]{@{}l@{}}4. standard deviation for randomized\\     expansion\end{tabular}                                                                  \\ \bottomrule
\end{tabular}
}
\end{table}

\subsection{Workload of a distributed entity}
Algorithm \ref{nodeworkload} shows the workload of a distributed node (hospital) where the actual data perturbation is conducted. As the algorithm shows, the data will not be transmitted to the central entity as part of the perturbation process. First, the local parameters which are determined based on the local dataset will be sent to the central node. Next, the global parameters which are calculated and sent to the distributed entities will be used for the perturbation of the local dataset. However, the step of randomized expansion (refer step \ref{alrandexp} of Algorithm \ref{nodeworkload}) doesn't involve any interactions with the central node as randomized expansion needs to evaluate the noise based on each individual instance. Consequently, in the distributed setting, randomized expansion increases the randomization of the perturbed data over the centralized approach.  

\begin{center}
    \scalebox{0.8}{
    \begin{minipage}{1.1\linewidth}
     \removelatexerror
      \begin{algorithm}[H]
\caption{Task of a distributed entity}\label{nodeworkload}
\linespread{1.5}

\KwIn{
\begin{tabular}{ l c l } 
		$D_i$, $M\theta$,  $TN^{noise}$, $RF_{optimal}$, $STDVEC$, \\ $MEANVEC$   
\end{tabular}
}

\KwOut{
\begin{tabular}{ l c l } 
		$D_i^{p}$ &  $\gets$ & perturbed data partition 
\end{tabular}
}
   
 generate $D_i^{N}$\;
 generate $Cov(D_i^{N})$\;
 generate $Mean(D_i^{N})$\;
 send Phase 1 data to the central entity\;
 receive Phase 2 data from the central entity\;

 $D_i^{pt}=(M\theta \times TN^{noise}\times RF_{optimal}\times (D_i^{N})^T)^T$ \label{al3step2122}\;
 $D_i^{pt}=D_i^{pt}+\mathcal{N}(0,\,\min(\sigma_{D_i^{pt}}))$\label{alrandexp}\;
 $D_i^{p}$=$\  D_i^{pt} \bullet STDVEC$+$MEANVEC$\;
 randomly swap the tuples of $\ D_i^{p} $\;

\end{algorithm}
    \end{minipage}%
    }
    
  \end{center}

\subsection{Workload of the central node}
Algorithm \ref{distributedmethod} shows the workload of the central entity. As shown in the algorithm, the central entity calculates the global perturbation parameters that need to be transmitted to the distributed entities.  We assume that the distributed nodes are selected through a prior handshake mechanism that confirms the validity and reliability of the distributed nodes. However, since the proposed algorithm does not share the original data with any party before the perturbation, there will not be any privacy leak during the communication of the parameters. 

\begin{center}
    \scalebox{0.8}{
    \begin{minipage}{1.1\linewidth}
     \removelatexerror
      \begin{algorithm}[H]
\caption{Task of the central entity}\label{distributedmethod}

\KwIn{
\begin{tabular}{l c l } 
		$\ Cov(D_1),Cov(D_2),...,Cov(D_k) \gets$ Covariance\\    matrices of the data partitions\\
		$\ Mean(D_1),Mean(D_2),...,Mean(D_k) \gets $ Mean \\vectors of the data partitions
\end{tabular}
}

\KwOut{
\begin{tabular}{ l c l } 
		$M\theta$,  $TN^{noise}$, $RF_{optimal}$, $STDVEC$,\\ $MEANVEC$
\end{tabular}
}
receive Phase 1 data from the distributed entiteis\;
 $\Phi=0$\;
 $\theta_{optimal} = 0$\;
 generate $\ TN^{noise} $\; 
 $Cov(D_t)=Cov(D_1)$\;
\For{\textbf{each} covariance matrix, $Cov(D_{i})$ }{
			\If{$i<k $}{
			 $Cov(D_o)=Cov(D_{i+1})$\;
			 $Cov(D_{new})=merge(Cov(D_t),Cov(D_o))$, according to Equation \ref{matcovupdate}\;
             $Cov(D_t)=Cov(D_{new})$\;
            }
}
 $Cov(D)=Cov(D_{new})$\;
 $STDVEC=\sqrt{diag(Cov(D))}$\;
 $MEANVEC= combine(Mean(D_1),Mean(D_2),...,Mean(D_k))$, using,
 $MEANVEC=\frac{m_1\times\overline{y_1}+m_2\times\overline{y_2}+\hdots+m_k\times\overline{y_k}}{m_1+,_2+\hdots +m_k}$,\\
 where, $y_i=Mean(D_i)$\;
 
\For{\textbf{each} $ax$ in  $\left\{1,2,\dots,n\right\}$}{  \label{al3step722}
			 generate $RF_{ax} $ \;
			\For{\textbf{each} $\theta_i $}{
			 generate $M_i $ using Algorithm \ref{rotmatgenerate}\;
             $\ \phi_{i,{ax}}=min(\vec 1+diag(M_i\times RF_{ax}\times\; Cov(D^{N})\times RF_{ax}\times M_i^T)+sum((Cov(D^{N})\times RF_{ax} \bullet M_i)^T)^T)$ \label{al2step1122}	
			}			
			} \label{al3step1822}
			\For{\textbf{each} $\theta_i $}{ 
               $\ \phi_{i}=min(\phi_{i,{ax}})$  where, $ax$ $\in$  $\left\{1,2,\dots,n\right\}$\ \label{step19al2};	
			}			
             $\Phi = max(\phi_{i})$ where, $i$ $\in$  $\left\{1,2,\dots,\theta \right\}$\;
             $\theta_{optimal}=\theta_i$  at $\Phi $\;
			 $Rif_{optimal}=ax$  at $\Phi $\;
            
			 generate $M\theta $ \;
			 generate $RF_{optimal} $ \;
send Phase 2 data to the distributed entities\;
\end{algorithm}
\end{minipage}%
}
\end{center}

\subsection{Federated learning over perturbed data}
As shown in Figure \ref{distributed}, the federated module comes into action after the perturbation is completed  (in the first two phases). Each of the distributed entities will use the local data (perturbed) to train a local ML model (e.g. deep neural network) for a certain number of local epochs. After finishing the local epochs, the distributed entities will send the model parameters to the central repository to generate a global representation of the model by aggregating the model parameters (refer to section \ref{fedlearning}). The server (central authority) then passes the aggregated parameters back to the distributed entities to update the local models to generalize the models. This is called one round of federation. The federated learning setup will conduct a sufficient number of local epochs and federation rounds to train the ML models based on the requirements of the organization (e.g. production of data streams or big data), ML model architecture, and the properties of the input dataset. 

\section{Results}
\label{resdiscussion}
This section provides the experimental results of the performance of DISTPAB. We tested DISTPAB on six datasets to compare and evaluate its performance against three algorithms: RP (rotation perturbation), GP (geometric perturbation), and PABIDOT. We considered $0.3$  as the default value of $\sigma$ for the experiments unless specified otherwise.  Next, we measured the performance of DISTPAB on FedML to examine the utility loss due to the perturbation. More details on the FedML setup is provided in section \ref{fedlearn}. For the experiments, we used a  Windows 7 (Enterprise 64-bit, Build 7601) computer with an Intel(R) i7-4790 (4$^{th}$ generation) CPU (8 cores, 3.60 GHz) and 8GB RAM. We declared a virtual distributed environment (refer Section \ref{parpool}) to run DISTPAB under the computer settings mentioned above. Table \ref{datasettb} has a summary of the datasets used for the experiments. We selected the datasets by considering a diverse range of domains. We first perturbed the data using  RP, GP, PABIDOT, and DISTPAB.  Next, we used the perturbed data to evaluate and compare the attack resistance and the classification accuracy of  RP, GP, PABIDOT, and DISTPAB. For classification accuracy analysis, we used Weka 3.6~\cite{witten2016data}, which is a data mining tool that packages a collection of data mining algorithms. We used the following classification algorithms: Naive Bayes, k-nearest neighbor (kNN)\footnote{In Weka,  kNN is named as IBk (Instance Based Learner)},  Support vector machine (SVM)\footnote{In Weka, Sequential Minimal Optimization (SMO) algorithm is used to train an SVM}, Multilayer perceptron (MLP), and J48 ~\cite{witten2016data} to investigate the utility of the perturbed data. 

\begin{table}[H]
\centering

    \caption{Information about the generic datasets used for the experiments. We selected the data based on varying dimensions from small to large in order to test the behavior of DISTPAB on different dynamics of the input data dimensions.}   
    \label{datasettb}

    \begin{small}
    	\setlength\tabcolsep{5pt} 
        \resizebox{1\columnwidth}{!}{
    \begin{tabular}{ l l l l l }
    \toprule
{\bfseries Dataset} & {\bfseries  Abbreviation}         & {\bfseries Number of Records}     & {\bfseries Number of Attributes }   &   \bfseries{Number of Classes}   \\
    \midrule
Wholesale customers\tablefootnote{https://archive.ics.uci.edu/ml/datasets/Wholesale+customers}       &	 WCDS & 440 \ & 8 \ & 2 	\\

    Wine Quality\tablefootnote{https://archive.ics.uci.edu/ml/datasets/Wine+Quality}      & WQDS & 4898 \ & 12 \ & 7 \\
    
     Page Blocks Classification \tablefootnote{https://archive.ics.uci.edu/ml/datasets/Page+Blocks+Classification}          & PBDS  & 5473 \ &  11 \ &   5  \\
     
Letter Recognition\tablefootnote{https://archive.ics.uci.edu/ml/datasets/Letter+Recognition}             &  LRDS	& 20000 & 17 & 26\\

       Statlog (Shuttle)\tablefootnote{https://archive.ics.uci.edu/ml/datasets/Statlog+\%28Shuttle\%29}        &  SSDS & 58000 \ & 9 \ & 7 \\

HEPMASS\tablefootnote{https://archive.ics.uci.edu/ml/datasets/HEPMASS\#}      & HPDS & 3310816  \ & 28 \ & 2 \\

    \bottomrule
    \end{tabular}
    }
    \end{small} 
    
\end{table}

\subsection{Distributed computing setup}
\label{parpool}
To test DISTPAB in a distributed setting, we used the ``parpool" function in MATLAB~\cite{leon2016controlling}. ``parpool(N)" creates N number of parallel processors (named as parallel workers) which can perform distributed computing. For performance testing and comparison, we distributed an input dataset among four workers (unless specified otherwise) by equally dividing the input dataset to make sure each distributed entity entails the same computational workload during the experiments. 

\subsection{Time Complexity}
\label{timecomplexitydistpab}
We used both theoretical and runtime analysis to evaluate the computational complexity of DISTPAB. First, we conducted theoretical evaluations of the computational complexity of DISTPAB. Next, we conducted runtime analyses to provide empirical evidence on the estimated time complexities. The section from line \ref{al3step722} to line \ref{al3step1822} (in Algorithm \ref{parallelalgo}) has two loops that have a significant effect on the computational complexity of Algorithm \ref{parallelalgo} as one is constrained by the number of attributes whereas the other is controlled by the range of rotation angle. We can estimate that the runtime complexity of the  corresponding loops is $O(k\times n)=O(n)$ ($n$ = the number of attributes, $m$ = the number of tuples, and $k$ is a constant for the constant range of angle).  We can estimate that step \ref{al2step11} (of Algorithm \ref{parallelalgo}) is $O(n^3)$. The loop segment (starts from step \ref{al3step7} to \ref{al3step18}) contributes a $O(n^4)$ complexity. In can also be estimated that  step \ref{al3step21} introduces the highest computational complexity of $O(n^3\times m)$ as $m>>>n$. In terms of worst-case computational complexity, we can state that Algorithm \ref{parallelalgo} has a computational complexity  $O(n^3\times m)$ when $m>>>n$. The steps  \ref{al3step7} to \ref{al3step18} of   Algorithm \ref{parallelalgo} are moved to Algorithm \ref{distributedmethod} which runs on the central node, and  step \ref{al3step21} of  Algorithm \ref{parallelalgo} is moved Algorithm \ref{nodeworkload} which runs on the distributed nodes. Consequently, the central node will have a computational complexity of $O(n^4)$. The distributed nodes will have a computational complexity of $O(n^3\times m)$. However, the communication delay can increase the data perturbation time in a distributed setting.

\begin{figure}[H] 
	\centering
	\subfloat[The time consumption by the central entity for perturbation against the number of attributes.  The plot confirms the theoretical evaluation of the central entity's time complexity, which is $O(n^4)$, where $n$ is the number of attributes.  ]{\includegraphics[width=0.48\textwidth, trim=0cm 0cm 0cm 0cm]{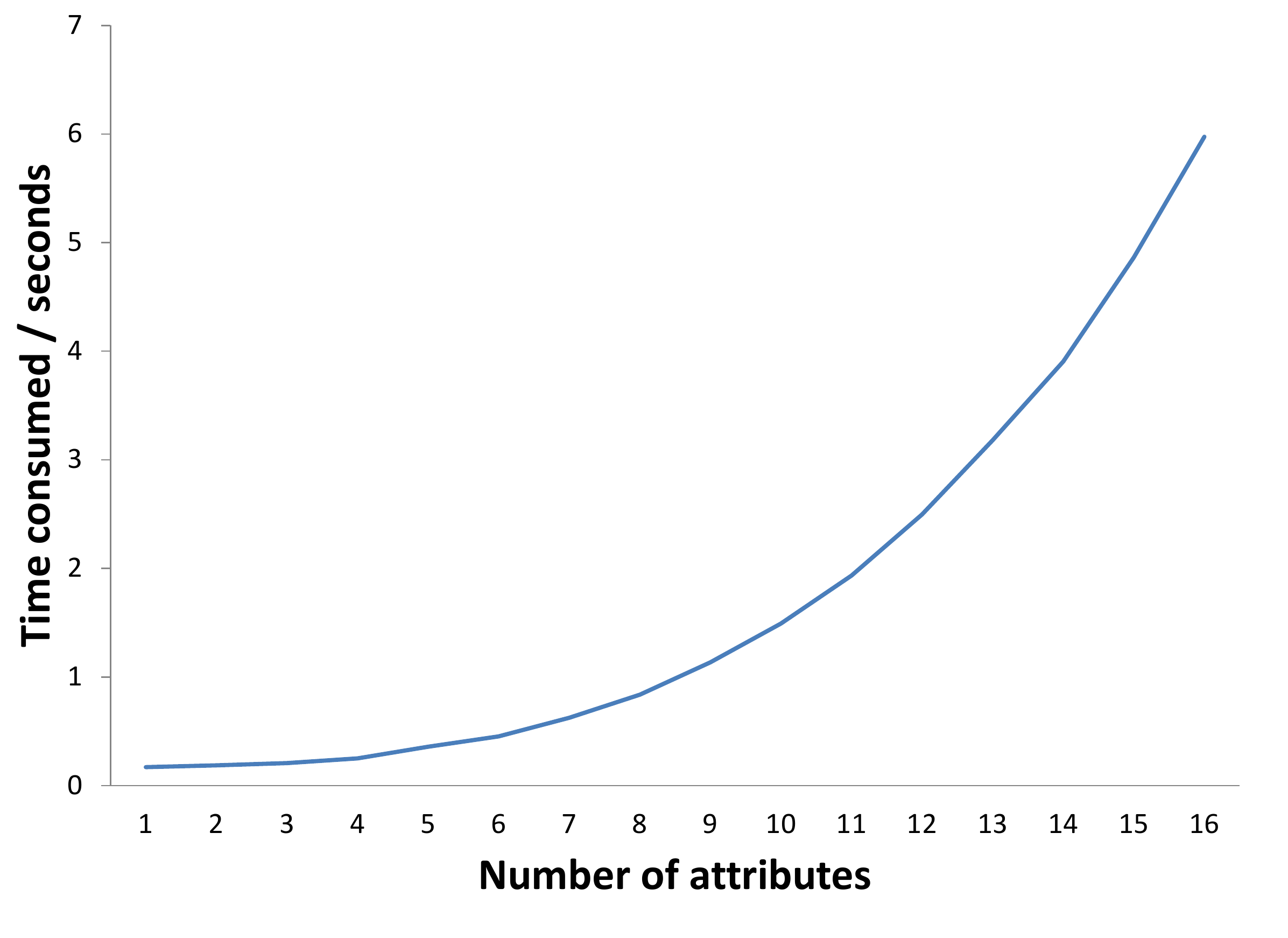}\label{central_att_time}}
	\hfill
	\subfloat[The time consumption by the central entity for perturbation against the number of tuples.  The plot confirms the theoretical evaluation of the central entity's time complexity, which is $O(n^4) = O(1)$ as $n$ is constant. The central node consumes a constant amount of time when $n$ (the number of attributes) is constant.  ]{\includegraphics[width=0.48\textwidth, trim=0.3cm 0cm 0cm 0cm]{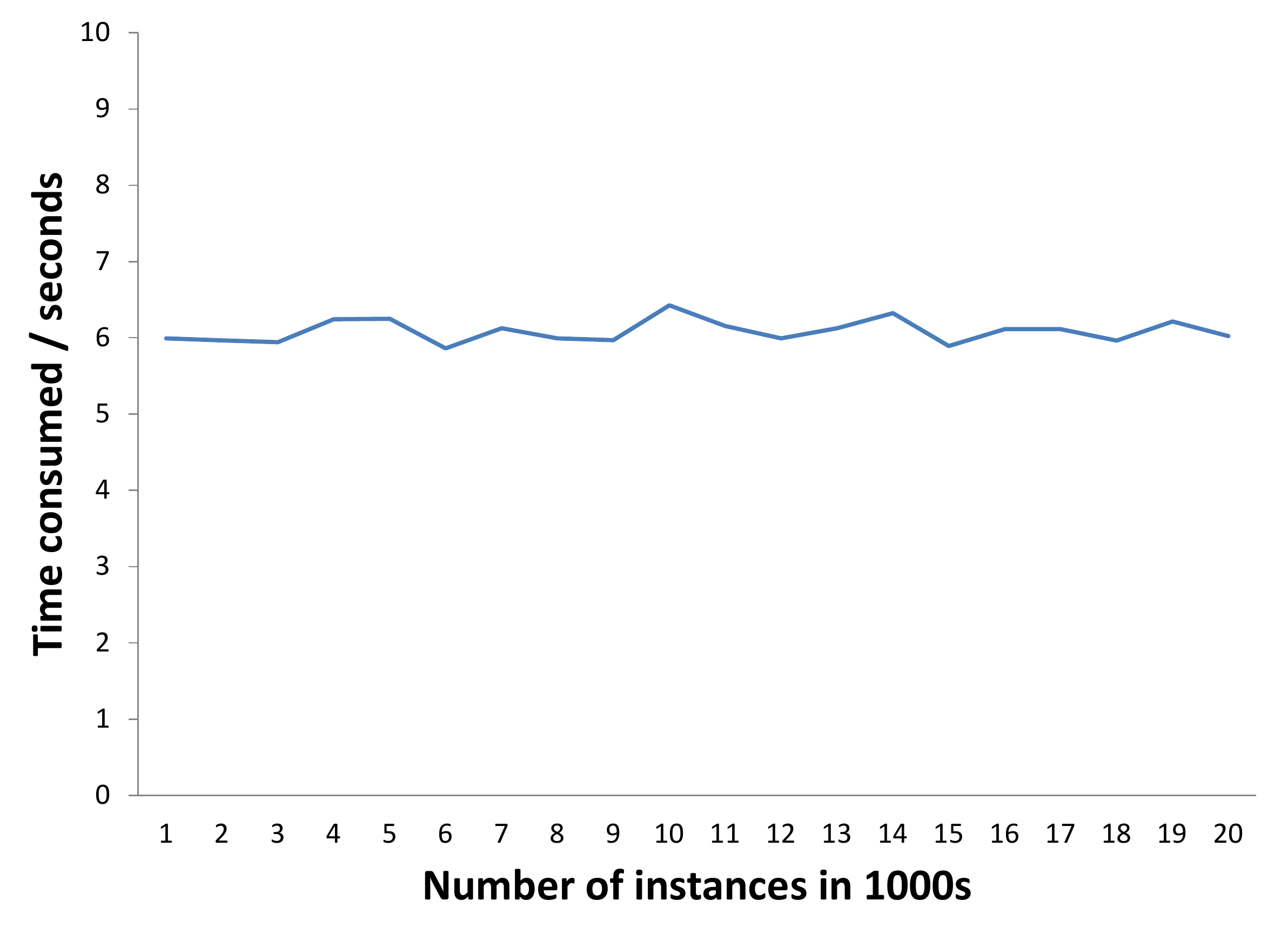}\label{central_att_instances}}

	\caption{Time consumption by the central entity. The plots do not include communication delays, and they show the time consumption for the perturbation of the LRDS dataset. }
    \label{centraltime}

\end{figure}

\begin{figure}[H] 
	\centering
	\subfloat[The average time consumption of the distributed nodes for the perturbation against the number of attributes.  According to the time complexity of $O(n^3\times m)$, as $m$  remains constant, the plot should show a time complexity of $O(n^3)$ where $m$ and $n$ represent the number of tuples and the number of attributes respectively.]{\includegraphics[width=0.48\textwidth, trim=0cm 0cm 0cm 0cm]{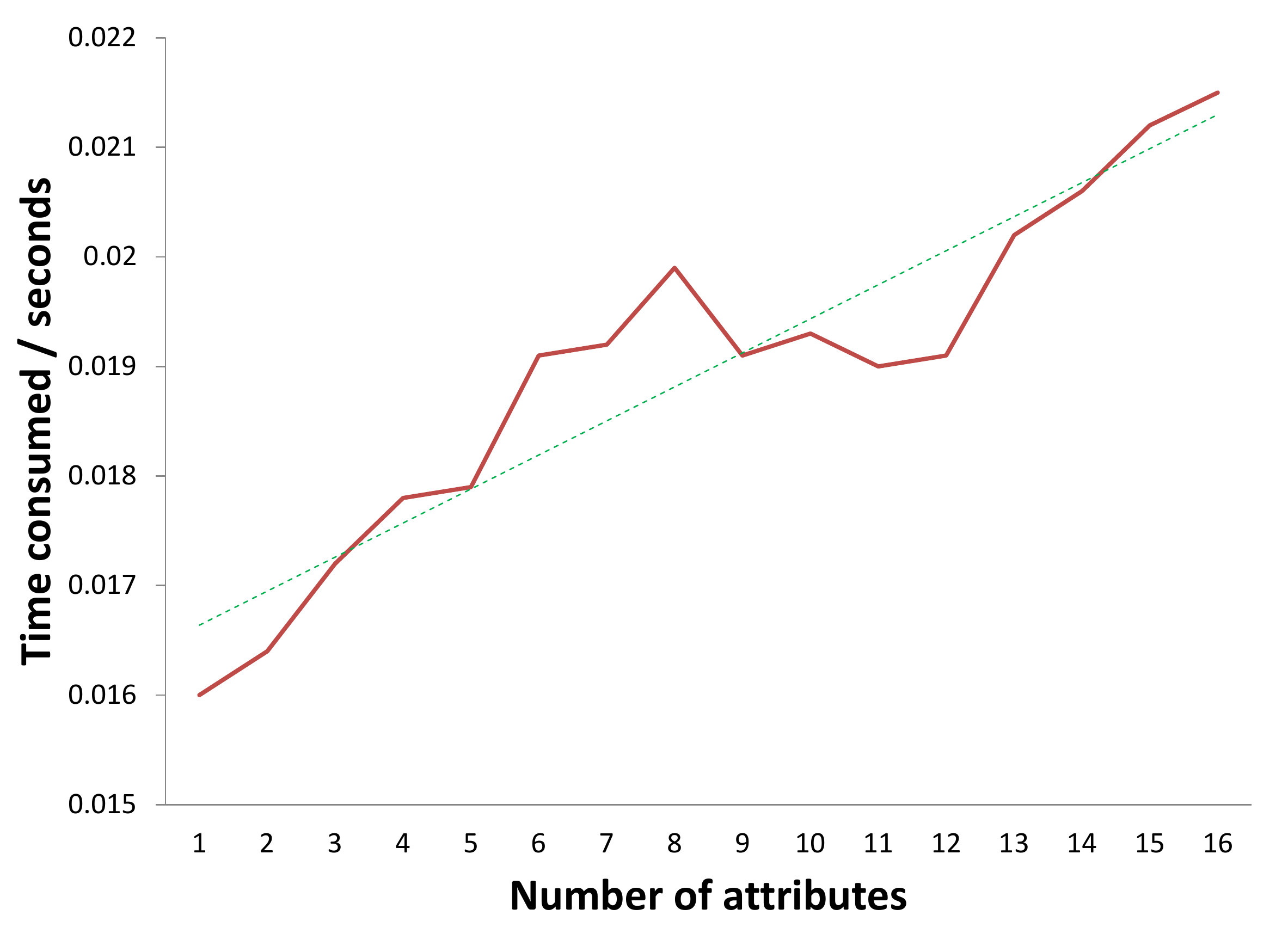}\label{distributed_att_time}}
	\hfill
	\subfloat[The average runtime consumption of the distributed nodes for the perturbation against the number of instances. The plot confirms the theoretical evaluation of the time complexity, which is $O(m)$, where $m$ is the number of tuples.]{\includegraphics[width=0.48\textwidth, trim=0.3cm 0cm 0cm 0cm]{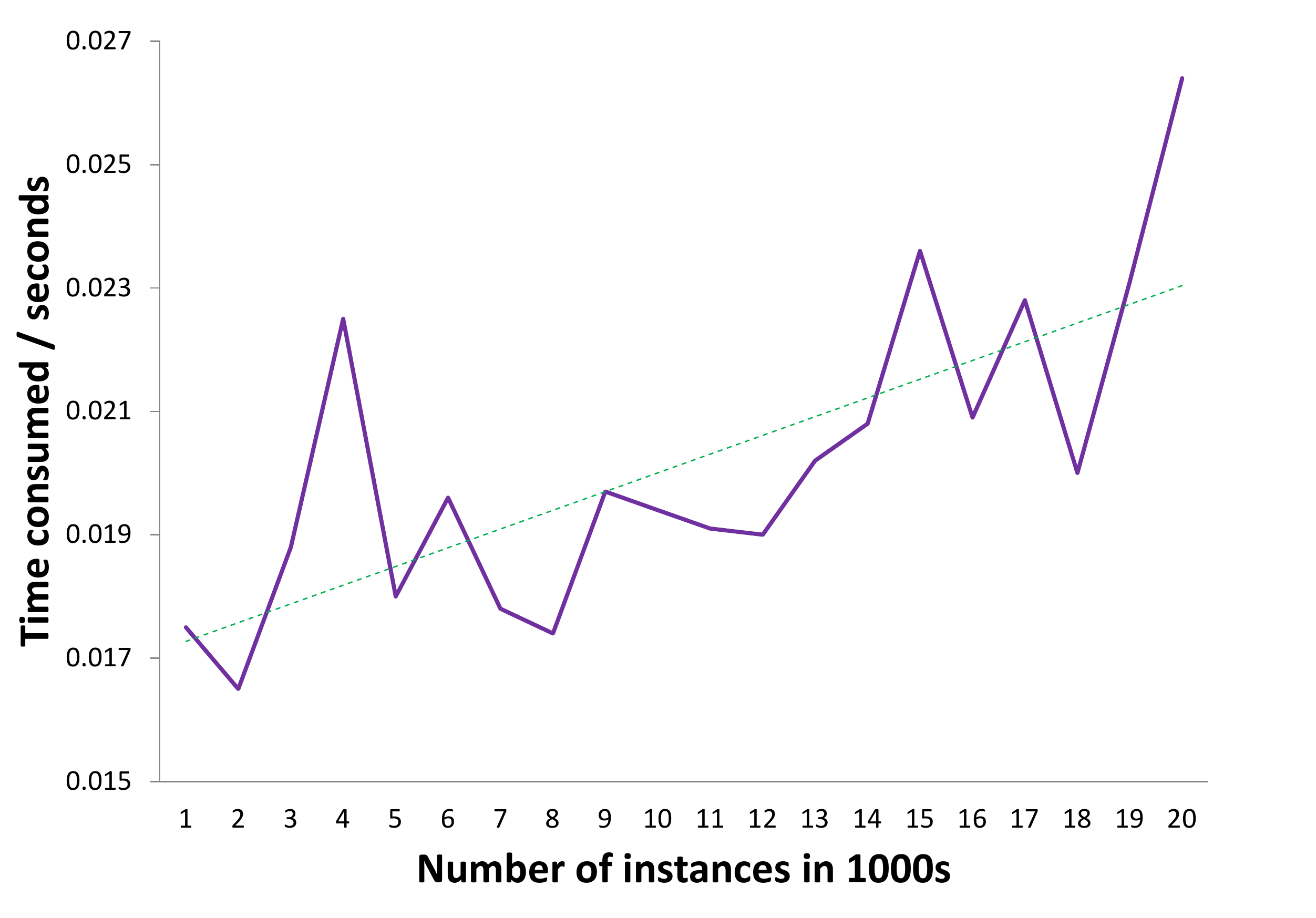}\label{distributed_att_instances}}

	\caption{Average time consumption by the central entity (for 4 nodes, refer Section \ref{parpool} for the specifications of the distributed setup).  The plots do not include communication delays, and they show the time consumption for the perturbation of the LRDS dataset.}
    \label{distributedtime}

\end{figure}

Figures \ref{centraltime} and \ref{distributedtime} show the empirical trends of time consumption by the central node and the distributed nodes in perturbing the LRDS dataset. During the time calculations, we considered only the time consumed for the perturbations by the corresponding processing unit (central or distributed) to get an absolute understanding of time consumption. We ignored all the other factors, such as communication delays and I/O operation delays. For this experiment, we considered four distributed nodes communicating with the central node (refer Section \ref{parpool}). We used the LRDS dataset for the analysis. During the study on how the time consumption is affected by the number of attributes, we distributed the LRDS dataset equally among the distributed nodes (each processing 5000 tuples). Next, we evaluated the effect of the number of distributed nodes on the classification accuracy and attack resistance by gradually increasing the number of distributed nodes from 1 to 20. For this experiment also, we used the LRDS dataset, and for each case, the dataset was distributed equally among the distributed nodes.
 
\subsection{Complexity of distributed data perturbation vs. centralized data perturbation} 
\label{distpabvspabidot}
In this section, we first compare the computational complexities of DISTPAB and PABIDOT (the centralized approach). Next, we investigate the impact of communication costs on data perturbation. As explained in Section \ref{timecomplexitydistpab}, in DISTPAB, the central node has a worst-case computational complexity of $O(n^4)$, while a distributed node has a worst-case computational complexity of $O(n^3\times m)$ (where $n$ is the number of attributes and $m$ is the number of instances). Since the number of attributes of a given setup is fixed and $m>>>n$, the computational complexity of DISTPAB is governed by  $O(n^3\times m)$. The worst-case computational complexity of PABIDOT is also $O(n^3\times m)$~\cite{chamikara2020efficient}. Consequently, for a given setup with a fixed number of attributes (a fixed number of IoT sensors), both DISTPAB and PABIDOT will provide linear complexity ($O(m)$). The only factor that would increase the time consumption of DISTPAB compared to PABIDOT is the data communication cost. 

\begin{figure}[H] 
	\centering
	\subfloat[Time consumption of the centralized version and DISTPAB against the number of attributes. Both algorithms show similar trends (exponential) for the number of attributes.]{\includegraphics[width=0.48\textwidth, trim=0cm 0cm 0cm 0cm]{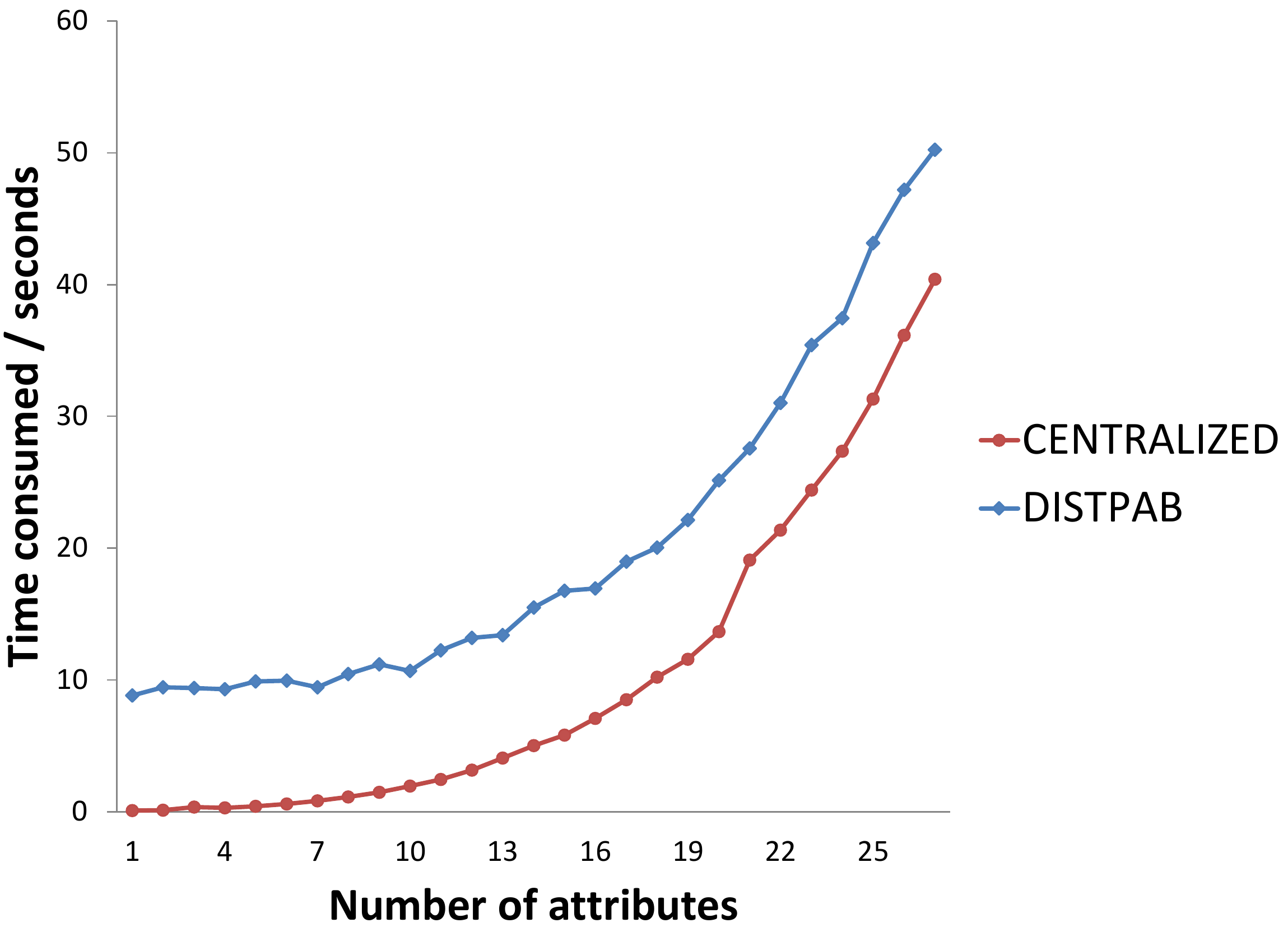}\label{distcompatt}}
	\hfill
	\subfloat[Time consumption of the centralized algorithm and DISTPAB against the number of tuples.The plots confirm the linear complexity of both versions for the number of tuples.]{\includegraphics[width=0.48\textwidth, trim=0.3cm 0cm 0cm 0cm]{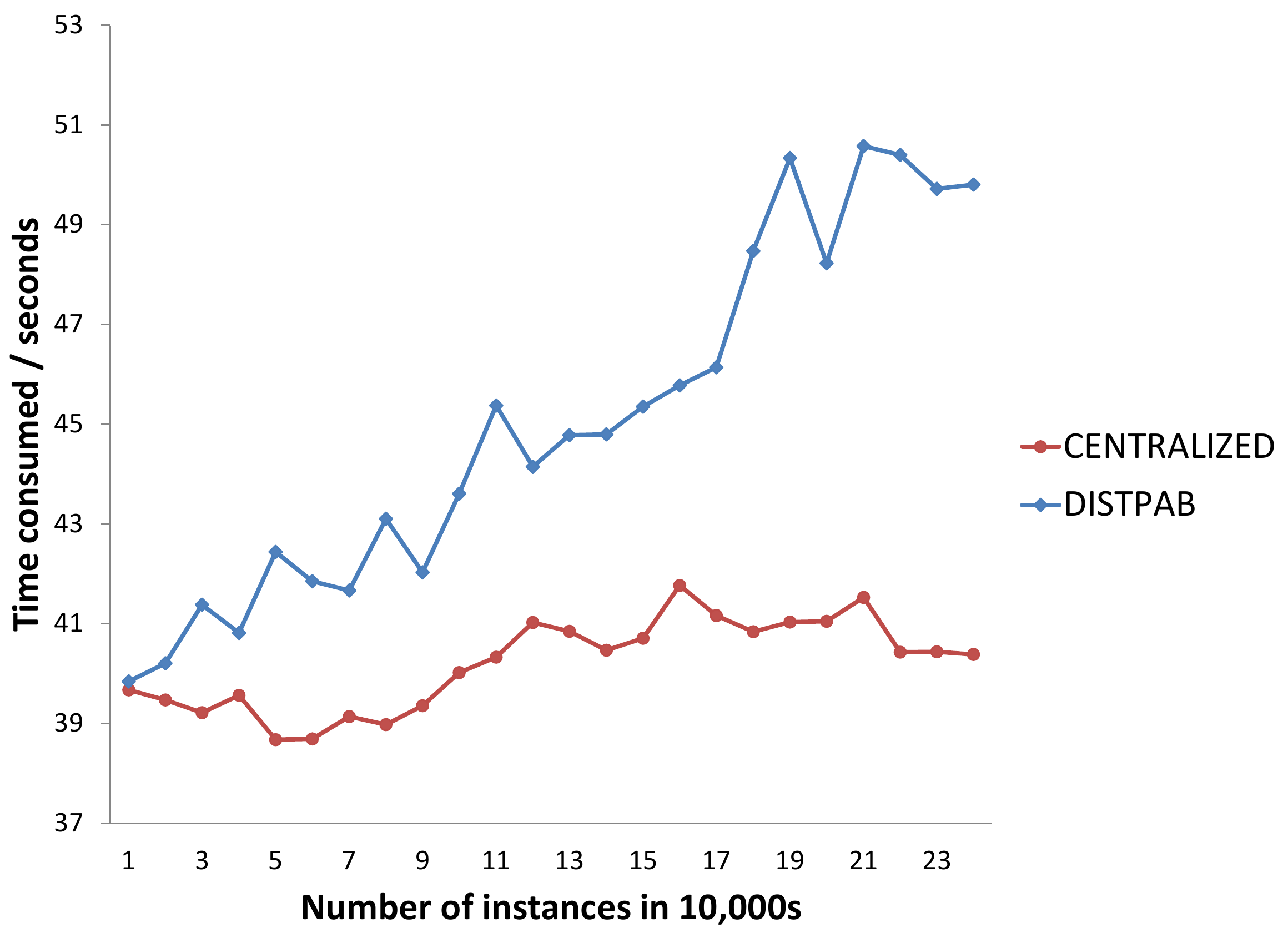}\label{distcompinst}}

	\caption{Impact of communication delay over distributed perturbation in DISTPAB.  In a real-world scenario, network communication can be influenced by many factors, such as network latency and network congestion. It is complex to get an exact estimation over the communication delays. As the figures show, DISTPAB follows the expected time complexity (as available in the plot for the centralized algorithm) while adding the communication delays to its patterns.}
    \label{distributedtimediff}
\end{figure}

As shown in Table \ref{datacom}, during the data perturbation process, DISTPAB transfers 3 parameters (a matrix of size $n\times n$, a vector of size $n$, and a number) from a distributed node to the central entity, while 4 parameters (three matrices of size $n\times n$, and a standard deviation value) from the central node to a distributed node, where $n$ is the number of attributes. Consequently, the total worst-case communication cost per node can be approximated as $O(4n^2+n+1)$, and for $p$ number of nodes, the worst-case computational complexity can be approximated as $O(p(4n^2+n+1))$, which is independent of the number of data instances (tuples). For a given setup, the values of $p$ and $n$ are constants. Consequently, the communication cost is constant, no matter how many data tuples is introduced to DISTPAB.  For empirical analysis of data communication costs, we use a large enough portion (240550$\times$28) of the HPDS dataset to obtain a significant enough trend while avoiding extensive time consumption. Figure \ref{distcompatt} shows the time consumption of the DISTPAB and the centralized version for an increasing number of attributes (the number of instances is kept constant). Figure \ref{distcompinst} shows the time consumption of DISTPAB for an increasing number of instances (while the number of attributes is kept constant). In both cases, DISTPAB consumes more time than the centralized version (PABIDOT) due to communication delays. However, according to the plots, the effect of communication delays on DISTPAB can be considered reasonable.

\subsection{Classification Accuracy}

\label{classaccusec}
\begin{figure}[H]
\centering
	\includegraphics[width=0.65\textwidth, trim=0.5cm 0cm 0.5cm 0cm]{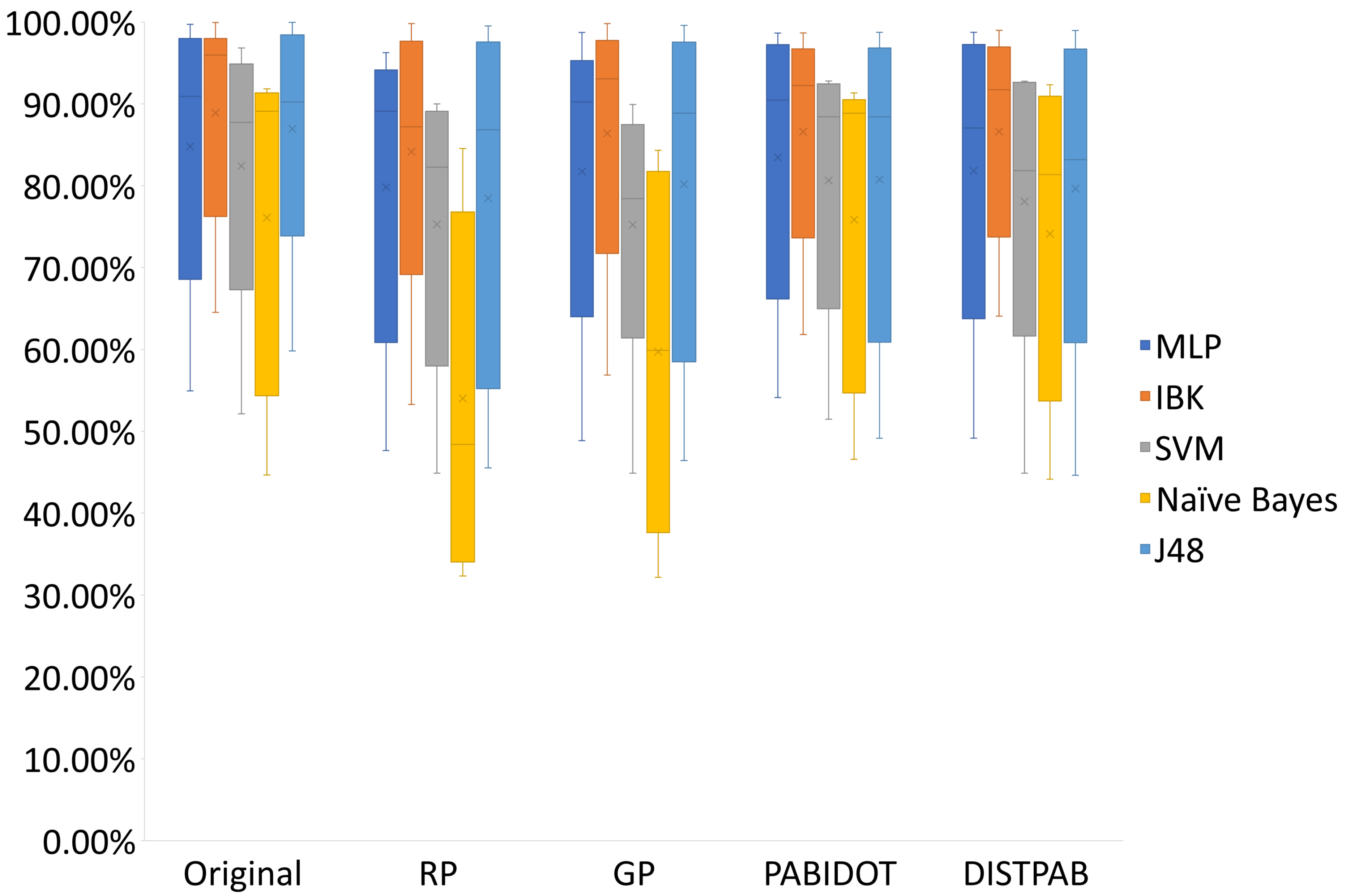}
	\caption{Classification Accuracy. The figure shows the variation of the classification accuracy produced by the original dataset and the datasets perturbed by the four perturbation approaches. In Weka,  kNN is referred to as IBk (Instance Based Learner).}
	\label{distributedaccuracy}
\end{figure}

Although the perturbation is distributed, DISTPAB always generates optimal global perturbation parameters similar to the perturbation parameter generation of the centralized algorithm (Algorithm \ref{parallelalgo}). However, the $randomized$ $expansion$ step is carried out by each distributed entity separately to improve the randomness of data. Consequently, the distributed version produces data with slightly higher randomization than its centralized version. This feature also reduces the utility of the data produced by the distributed version compared to the centralized version. Figure \ref{distributedaccuracy} shows the box plots of average accuracies produced by each of the algorithms against five classification algorithms. The classification accuracy was generated using 10-fold cross-validation. For the experiments on classification accuracy of a particular perturbed dataset, both classification model training and model testing were done using the same perturbed dataset. This is to check the model performance under strict privacy conditions where both the trained model and testing data do not leak any privacy to a third party.  As shown in the figure, DISTPAB and PABIDOT produce similar classification accuracies. The figure also shows that both DISTPAB and PABIDOT outperform RP and GP.

\subsubsection{Federated learning setup}
\label{fedlearn}
We simulated the federated learning setup using the Python socket programming interface and the \_thread interface.  In the default configuration, we considered 4 distributed clients. For this experiment, we used the SSDS dataset that has a sufficient number of tuples to see a noticeable impact on the performance while making a low burden on the computer resources (mentioned at the beginning of section \ref{resdiscussion}). We used a train/test split of 75\%/25\% during the experiments. We chose a simpler ANN model architecture for local models to avoid overfitting and achieve high efficiency with better training and testing accuracy. We estimated the number of hidden neurons to have a good model but at the same time to avoid overfitting by having too many neurons; the values were chosen after considering the data at hand. Each client trained a fully connected neural network; activation=`relu', batch size=64, hidden layer sizes=(10, 200, 200),  final layer size = num of classes, final layer activation = 'softmax', shuffle=True, optimizer=`SGD', learning rate=`constant', learning rate init=0.0001, momentum=0.5, verbose=True.  The models were trained for 20 federation rounds, while each model was locally trained for 3 epochs. However, it is essential to note that, for a different dataset, a different ANN architecture might need to be selected, as the performance of a particular ANN architecture is dependent upon the properties of the input dataset.

\begin{figure}[H]
\centering
	\includegraphics[width=0.5\textwidth, trim=0.5cm 0cm 0.5cm 0cm]{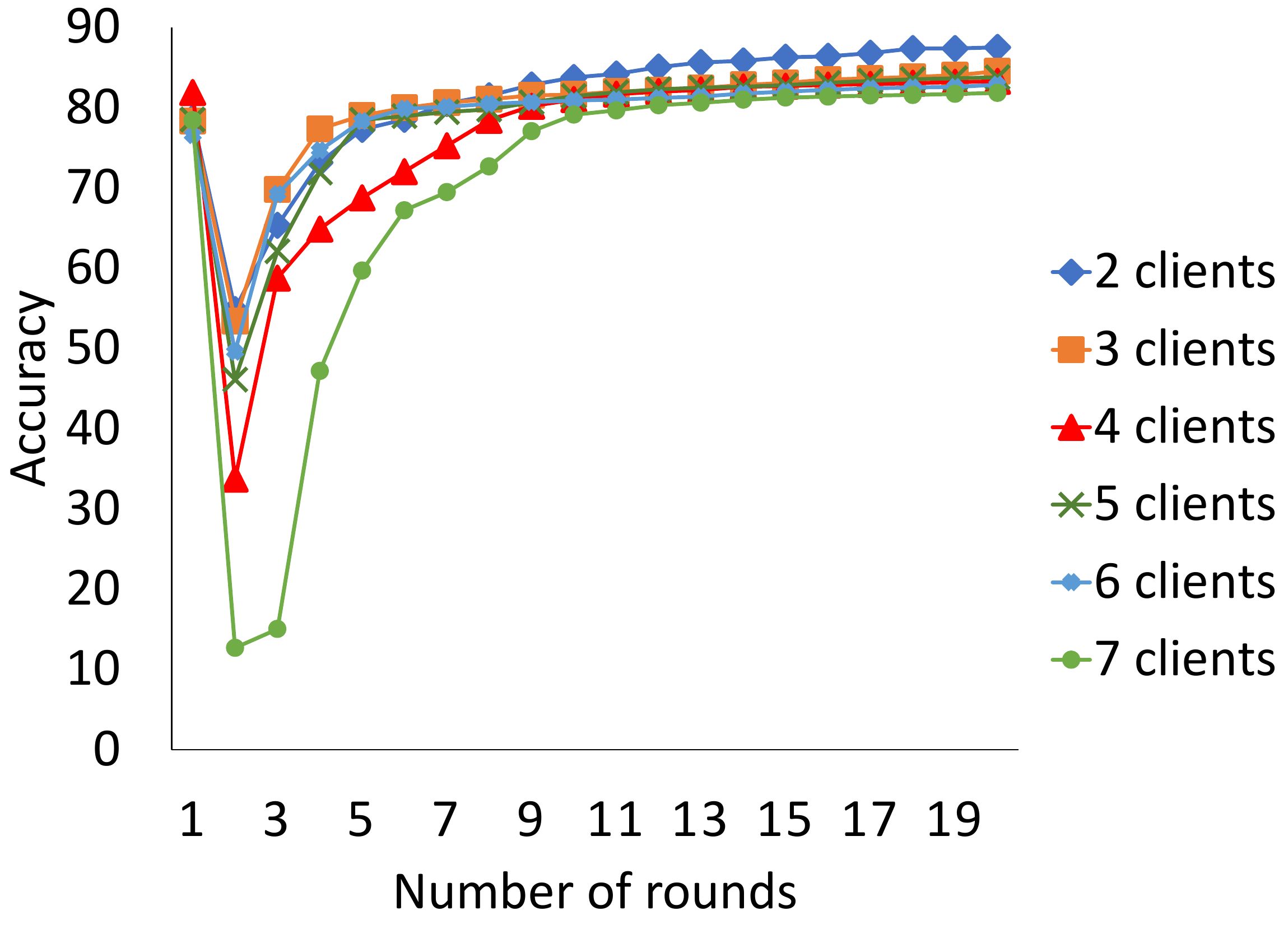}
	\caption{Classification accuracy during federated learning. The plots show the performance of federated learning on DISTPAB perturbed data under different numbers of distributed clients. The higher the number of clients, the higher the amount of time taken for ML model convergence. }
	\label{fedaccuracy}
\end{figure}

Fig \ref{fedaccuracy} shows the performance of the federated model using data perturbed by DISTPAB. The figure shows the performance against different number of distributed branches (clients). During the experiment, the input dataset was equally divided between distributed entities. Consequently, the higher the number of clients, the lower the number of tuples in each client. Thus, the model needs more federation rounds for  convergence when the number of clients is higher. We can also notice that there is a sudden drop in accuracy in the second round, and the accuracy increases as the federation continues. This is due to the extensive parameter modification that takes place during the second round. The higher the number of clients, the more significant the effect that results in a more substantial drop in the accuracy.

In the proposed setting, DISTPAB works as a stand-alone data perturbation mechanism (a local data perturbation approach), which needs to be applied to the raw input data before subjected to federated learning. Hence, a model that would not converge on non-perturbed (raw/original) data could not be considered for convergence under perturbed data. From an empirical standpoint, according to Figure \ref{distributedaccuracy}, we can notice that the classification accuracy generated by the perturbed data is very close to that of the original data. This implies that the perturbation employed by DISTPAB does not adversely affect the data representation. Consequently, we can assume that a model that can converge on the original raw data can converge on the perturbed data, subject to suitable hyperparameter tuning. However, more theoretical analysis is necessary to identify the exact effect of the perturbation on model convergence. The theoretical proof of convergence is beyond the scope of this paper~\cite{wang2019adaptive,dinh2019federated}. We consider this as a future direction of the proposed work.

\subsection{Scalability}
The scalability of the proposed approach is based on three factors: (1) the increase in the number of instances (number of tuples), (2) the increase in the number of attributes, and (3) the increase in the number of distributed entities. According to the computational complexity analysis (refer Section \ref{timecomplexitydistpab}), DISTPAB provides a linear complexity towards an increasing number of instances (for a given setup where the number of sensors (attributes) is fixed). Hence, for a given setup where the number of attributes and the number of distributed entities are fixed, DISTPAB will provide high scalability. When the number of sensors of the environment (which corresponds to the number of attributes of the dataset) is increased, the worst-case computational complexity changes to $O(n^4)$, where $n$ is the number of attributes. However, the number of attributes would not increase as frequently as the number of instances, and this increase is much smaller compared to the instances. When it comes to the distributed entities, the number of parameter communications will increase with the number of distributed entities. Compared to the number of instances, and the number of attributes, the increase in the number of distributed entities is extremely small. Hence, the increase in the number of distributed entities will not have a drastic impact on the scalability of DISTPAB. Hence, we can consider DISTPAB to provide high scalability towards distributed data perturbation and machine learning.

\subsection{Effect of the distribution of data instances (tuples) among distributed entities on the performance of perturbation}

A distributed entity can hold a certain number of data instances that match the corresponding computing node's capacity. In a conventional distributed setup, the distributed entities can have different computational resources and hold different numbers of data instances. It is important to investigate how this asymmetry can play a role in the performance of data perturbation employed by DISTPAB. Different distributed entities with different numbers of instances will consume different amounts of perturbation time. For a given distributed setup, a distributed entity has a data perturbation complexity of $O(m)$ (refer to Section \ref{timecomplexitydistpab}).  Besides, as discussed in Section \ref{distpabvspabidot}, when the number of attributes is fixed, DISTPAB exhibits an overall complexity (computation + communication) of $O(m)$. Hence, the differences in the numbers of instances in distributed entities will not have an adverse effect on the overall complexity of DISTPAB. As explained in Section \ref{ourapproach}, before the application of perturbation, DISTPAB generates the global perturbation parameters based on the local perturbation parameters retrieved from the distributed entities. As a result of this step, the perturbation effect applied by DISTPAB is almost the same as the effect of perturbation applied by the centralized approach (PABIDOT). As shown in Figure \ref{distributedaccuracy}, DISTPAB ends up producing classification accuracy similar to that of PABIDOT. Hence, there is no effect of the number of instances in each distributed entity on the classification accuracy.

\subsection{Attack Resistance}
\label{attres}
\begin{figure}[H]
\centering
	\includegraphics[width=0.5\textwidth, trim=0.5cm 0cm 0.5cm 0cm]{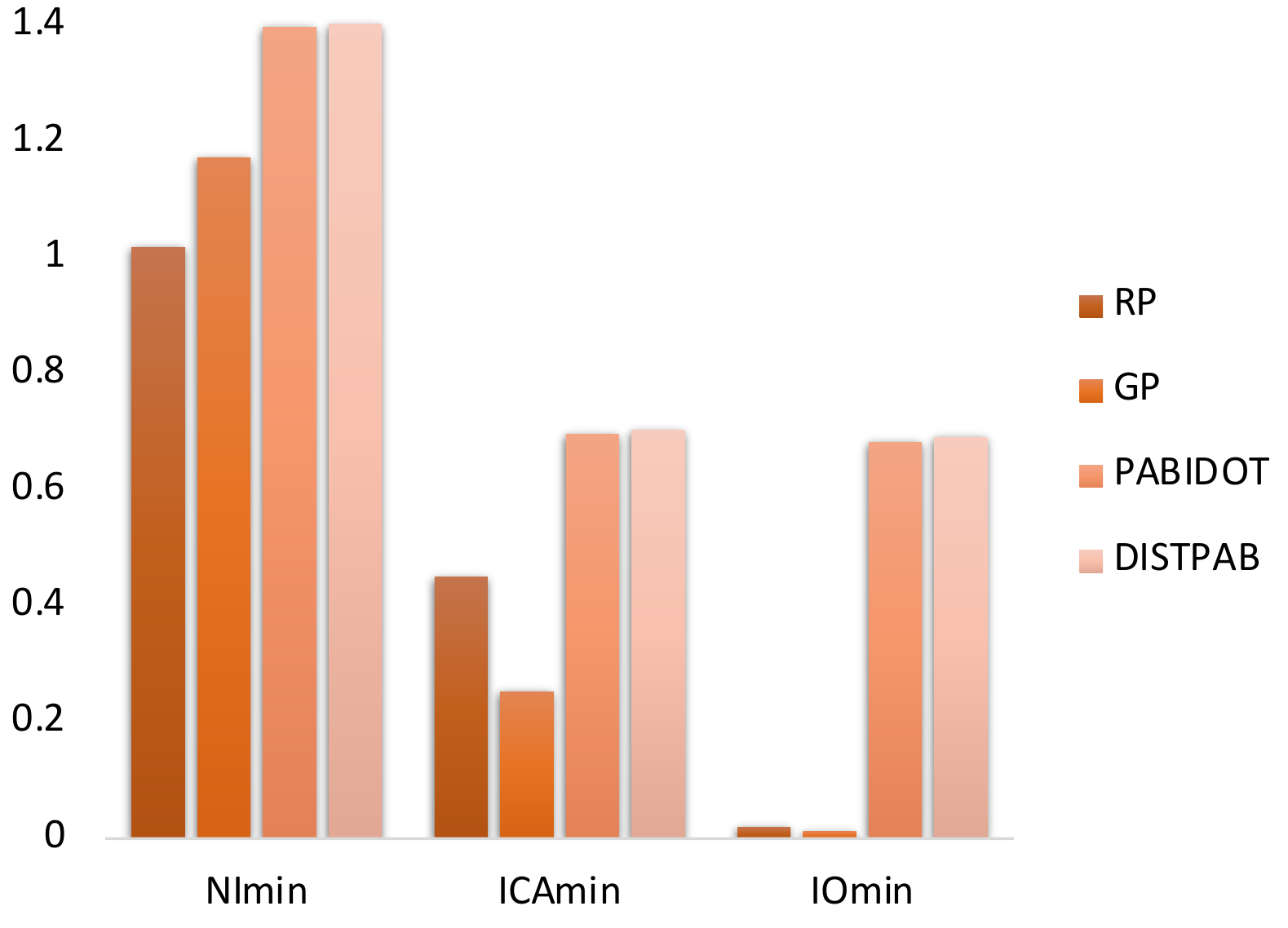}
	\caption{Attack Resistance. The figure shows the plots of the minimum values (NImin, ICAmin, and IOmin) of the $std (D-D^p$) under NI, ICA, and IO attacks. The higher the bar, the better the resistance to attacks.}
	\label{resiliencebar}
\end{figure}

We investigated the attack resistance of DISTPAB against Known I/O (IO) attacks, naive inference (NI), and Independent Component Analysis (ICA) ~\cite{okkalioglu2015survey} which are considered to be three potential attacks on matrix multiplicative approaches. We considered the default values of ten iterations and a sigma (noise factor) of 0.3  for the experiments on GP and RP.  We obtained $std (D-D^p$)) (std represents the standard deviation, $D$ the input data matrix, $D^p$ represents the perturbed data of D), which provides an estimate to the resistance against NI. Next, we applied ICA and IO on the perturbed data to generate reconstructed data. We assumed that 10\% of the original data as the background knowledge of an adversary during the IO attack investigation.  We obtained the minimum values (NImin, ICAmin, and IOmin) of the $std (D-D^p$) under each attack type and plotted them in Figure \ref{resiliencebar}. As shown in the figure, both PABIDOT and DISTPAB provide better attack resistance compared to RP and GP. We can see a slight increment in the minimum attack resistance of DISTPAB compared to PABIDOT due to the enhanced randomization as the $randomized$ $expansion$ step is carried out by each distributed entity separately to improve the randomness of data.

\section{Discussion}
\label{discussion}

This paper proposed an efficient distributed privacy preservation mechanism (named as DISTPAB) for distributed machine learning.  DISTPAB applies randomized $n-dimensional$ geometric transformations followed by randomized expansion, which is a noise application mechanism that improves the positiveness and negativeness of input data while improving randomization (used to further improve the randomization without harming the utility~\cite{chamikara2020efficient}).  DISTPAB uses $\Phi-separation$~\cite{chamikara2020efficient} as the underlying privacy model to obtain the optimal perturbation parameters and generate optimal privacy for the input data. We tested and compared the performance of DISTPAB against PABIDOT, RP, and GP for the classification accuracy, time complexity, and attack resistance.  Additionally, we investigated the impact of communication delay over distributed perturbation in DISTPAB, and further analyzed the impact of the number of distributed nodes on the classification during federated machine learning. 

According to our time complexity analysis, DISTPAB introduces a time complexity of $O(n^4)$ to the central entity, where $n$ represents the number of attributes. However,  $O(n^4)=O(1)$, as $n$ remains a constant for a given setting. Therefore, the central entity would consume a constant amount of time, no matter how many instances (tuples) are introduced.  A distributed entity shows a time complexity of $O(n^3\times m)$ = $O(m)$ as $n$ (the number of attributes) is a constant for a given dataset ($m$ represents the number of tuples). When we consider a fixed distributed setup, new parameters/sensors are rarely added at a speed in which the data grows; and more probably it will remain constant.  For a fixed distributed setting with a fixed number of distributed entities, the communication cost shows a constant complexity ($O(1)$). Consequently, for a given scenario, the amount of time consumed by a distributed node will grow linearly, which is optimal for privacy-preserving distributed machine learning.

The empirical evidence on data classification shows that DISTPAB provides similar classification accuracy to the centralized algorithm and better performance compared to RP and GP.  DISTPAB provides slightly lower classification accuracy compared to the centralized approach as DISTPAB imposes an increased level of randomization as the $randomized$ $expansion$ step is carried out by each distributed entity separately to improve the randomness of data. The empirical evidence (refer to Figure \ref{fedaccuracy}) proved that the number of distributed clients doesn't have a noticeable impact on classification accuracy during federated learning (distributed machine learning). However, the amount of time necessary for  model convergence increases with the number of clients.  This feature makes DISTPAB be the perfect solution for privacy-preserving federated learning (privacy-preserving distributed machine learning) where there is a large number of distributed entities involved. 

DISTPAB uses $\Phi-separation$~\cite{chamikara2020efficient} as its underlying privacy model, which allows optimal data perturbation for a given instance. Geometric data transformations and randomized expansion noise addition followed by random shuffling allow DISTPAB to impose high privacy by reducing the probability of data reconstruction attacks. Reverse z-score normalization of the final dataset takes the data back to their original attribute value ranges, making the attackers unable to distinguish the original data from the perturbed data. This feature reduces the chance of success of an attack trying to reconstruct the original input data from a perturbed dataset. DISTPAB provides better attack resistance compared to RP and GP. We can observe that the DISTPAB provides slightly better attack resistance compared to PABIDOT because DISTPAB adds an increased level of randomization as the randomized expansion step is carried out by each distributed entity separately to improve the randomness of data. However, in a federated learning setup, we do not share the perturbed data among the distributed clients or the server. Consequently, there will not be any data reconstruction attacks on the perturbed data, and due to the strong notion of privacy of the perturbed data, the privacy of the trained models will also be high.

In essence, DISTPAB can be an optimal privacy preservation solution for data privacy and machine learning that control extensive amounts of data ~\cite{manogaran2017big}, which are deployed in geographically distributed systems.

\section{Conclusions}
\label{conclusion}

Many modern systems, such as healthcare and open banking, are often geographically distributed and constrained with proper mechanisms for privacy-preserving data sharing for analytics.  This paper proposed a distributed perturbation algorithm named DISTPAB that can enforce privacy for distributed machine learning.  In the proposed setup of DISTPAB, a central/coordinating entity controls the global perturbation parameter generation, whereas the distributed entities can conduct local data perturbation.  The computational complexity of the algorithm that runs in the central entity is $O(1)$ (constant time) for the number of instances. The computational complexity of the algorithm, which runs on a distributed entity, is $O(m)$ for the number of tuples as the number of attributes often remains constant in a given setting where $m$ is the number of instances.  Consequently, the operations on distributed entities have a low computational complexity resulting in excellent efficiency.  DISTPAB provides high classification accuracy, which is close to that of the accuracy of classification performed with the original data.  DISTPAB provides high attack resistance outperforming rotation perturbation and geometric perturbation.  It was also shown that the data produced by DISTPAB is not subjected to utility degradation, against the number of distributed entities. DISTPAB can be an excellent privacy preservation algorithm for distributed machine learning.

As future work, we are interested in looking at improving the efficiency of the proposed work for the number of attributes. To achieve this, we will investigate on the vertical federated learning scheme where the distributed clients have different feature spaces. Vertical federated learning allows dividing a particular dataset into partitions by the attributes leaving a specific client to work only on a fewer number of attributes, which can improve efficiency.

 \appendix

\subsection{n-Dimensional translation matrix generation}
\label{apptrans}
Equation \ref{translationmat} represents the $\ \textit{n-dimensional} $ homogeneous translation matrix $\ T_{ND} $ ~\cite{jones2012computer}. The translational coefficients $(t_{i(n+1)},\dots,t_{(n)(n+1)})$ are drawn from a uniform random distribution which  is bounded within $(0,1)$. The uniform random noise is restricted to $0< t_{i(n+1)} <1 $ as the input dataset's attribute standard deviations and means become 1 and 0 after the z-score normalization.

\begin{equation}
    T_{ND}=
    \begin{bmatrix}
    1 & 0 & 0 & \dots & 0  & t_{1(n+1)} \\
    0 & 1 & 0 & \dots & 0 & t_{2(n+1)} \\
    0 & 0 & 1 & \dots & 0  & t_{3(n+1)} \\
    \vdots & \vdots & & \ddots & & \vdots \\
    0 & 0 & 0 & \dots & 1  & t_{(n)(n+1)} \\
    0 & 0 & 0 & \dots & 0 & 1 \\
    \end{bmatrix}_{(n+1)\times (n+1)}
\label{translationmat}
\end{equation}

\subsection{n-Dimensional reflection matrix}
\label{appreflect}
Equation \ref{reflectionmat} shows the homogeneous reflection matrix $\ RF_{ND} $. Equation \ref{reflectionmat} represents the reflection across axis one ~\cite{jones2012computer}. The (n+1) axis reflection for the matrix given in  Equation \ref{reflectionmat} can be written as shown in Equation \ref{refmatn1}, which provides a low level of bias in the perturbation. 

\begin{equation}
    RF_{ND}=
    \begin{bmatrix}
    1 & 0 & 0 & \dots & 0 & 0 & 0 \\
    0 & -1 & 0 & \dots & 0 & 0 & 0 \\
    0 & 0 & -1 & \dots & 0 & 0 & 0 \\
    \vdots & \vdots & & \ddots & & \vdots & \vdots \\ 
    0 & 0 & 0 & \dots & -1 & 0 & 0 \\
    0 & 0 & 0 & \dots & 0 & -1 & 0 \\
    0 & 0 & 0 & \dots & 0 & 0 & 1 \\
    \end{bmatrix}_{(n+1)\times (n+1)}
\label{reflectionmat}
\end{equation}

Equation \ref{refmatn1} represents The (n+1) axis reflection matrix  matrix.

\begin{equation}
RF_{\overline{ND}}=
\begin{bmatrix}
-1 & 0 & 0 & \dots & 0 & 0 & 0 \\
0 & 1 & 0 & \dots & 0 & 0 & 0 \\
0 & 0 & 1 & \dots & 0 & 0 & 0 \\
\vdots & \vdots & & \ddots & & \vdots & \vdots \\ 
0 & 0 & 0 & \dots & 1 & 0 & 0 \\
0 & 0 & 0 & \dots & 0 & 1 & 0 \\
0 & 0 & 0 & \dots & 0 & 0 & 1 \\
\end{bmatrix}_{(n+1)\times (n+1)}
\label{refmatn1}
\end{equation}

\subsection{Generating the rotational matrix}
\label{approtate}
Algorithm \ref{rotmatgenerate} provides the steps in generating the rotational matrix using the concept of concatenated subplane rotation. To represent the entire $\ n-dimensional $ orientation, we use the concept of the concatenated sub-plane rotation. The block matrix given in Equation \ref{rotationmat}, shows the rotations on the plane represented by a pair of coordinate axes $\ (\hat{x_i}, \hat{x_j})$ for $1\leq i,j \leq n $ ~\cite{paeth2014graphics}.   Thus $\ \frac{N(N-1)}{2} $ distinct $\ R_{ij}(\theta_{ij}) $ should be concatenated in a particular order to generate the composite \textit{n-dimensional} orthonormal matrix as represented in Equation \ref{rotmul}  with $\ \frac{N(N-1)}{2} $ degrees of freedom, parameterized by$\ \theta_{ij} $. We generate $M$ , the \textit{n-dimensional} concatenated subplane rotation matrix for an angle, $\theta_i$ using Algorithm \ref{rotmatgenerate}  (in \ref{compalgosec}) for each $\theta_i$ where ($(0<\theta_i<\pi),\quad\theta_i \notin \left\{ \pi/6, \pi/4, \pi/3, \pi/2, 2\pi/3, 3\pi/4, 5\pi/6  \right\} $)~\cite{chamikara2020efficient}. 

\begin{equation}
    R_{ij}(\theta_{ij})=
    \begin{bmatrix}
    1 & \dots & 0 & 0 & \dots & 0 & 0 & \dots & 0 \\
    \vdots & \ddots & \vdots & \vdots & \ddots & \vdots & \vdots & \ddots & \vdots \\
    0 & \dots & cos \theta_{ij} & 0 & \dots & 0 & -sin \theta_{ij} & \dots & 0\\
    0 & \dots & 0 & 1 & \dots & 0 & 0 & \dots & 0 \\
    \vdots & \ddots & \vdots & \vdots & \ddots & \vdots & \vdots & \ddots & \vdots \\
    0 & \dots & 0 & 0 & \dots & 1 & 0 & \dots & 0 \\
    0 & \dots & sin \theta_{ij}& 0 & \dots & 0 &  cos \theta_{ij}  & \dots & 0\\
    \vdots & \ddots & \vdots & \vdots & \ddots & \vdots & \vdots & \ddots & \vdots \\
    0 & \dots & 0 & 0 & \dots & 0 & 0 & \dots & 1 
    \end{bmatrix}_{(n+1)\times (n+1)}
\label{rotationmat}
\end{equation}

\begin{equation}
M=\prod_{i<j} R_{ij}(\theta_{ij})
\label{rotmul}
\end{equation}

\label{compalgosec}
\begin{center}
    \scalebox{0.8}{
    \begin{minipage}{1.1\linewidth}
     \removelatexerror
      \begin{algorithm}[H]
	\caption{Rotation matrix generation}\label{rotmatgenerate}
			\KwIn{
			\begin{tabular}{l c l} 
				$n               $ & $\gets $ & number of attributes of the input dataset\\
				$\theta $ & $ \gets $ & angle of rotation \\
				
			\end{tabular}
			}
			\KwOut{
			\begin{tabular}{ l c l } 
				$M $ & $\ \gets $ & multidimensional rotation matrix of $ \theta $  
			\end{tabular}
			}
			 $V $ = $ \{1,2,3\hdots n\} $\;
			 $C$ = $ \{(i,j)| i,j \in V $ and $ i \neq j \}  $\;
				$I_n $ = identity matrix of size $n$ \;
				$N $ = $ \binom{n}{2} $\;
             $I3=I_n $\;
			\For{\texttt{ $ k = 1 $ $ to $ $ N $ }}{
    			 $ A=\{(i_k,i_k),(j_k,i_k),(i_k,j_k),(j_k,j_k)\} $ where $ \{i_k,j_k \}$  
                 is the $ k^{th} set$ $of $ $ C$\;
    			 $I2$ $= $ $I_{n}$\;
    			 $I2(i_{k1},j_{k1})$ $= $ $cos(\theta)$ 
    			 where, $(i_{k1},j_{k1})$ is the set number 1 of $ A $\;
    			 $I2(i_{k2},j_{k2})$ $= $ $sin(\theta)$ 
     			 where, $(i_{k2},j_{k2})$ is the set number 2 of $ A $\;
    			 $I2(i_{k3},j_{k3})$ $= $ $-sin(\theta)$ 
     			 where, $(i_{k3},j_{k3})$ is the set number 3 of $ A $\;
    			 $I2(i_{k4},j_{k4})$ $= $ $cos(\theta)$ 
     			 where, $(i_{k4},j_{k4})$ is the set number 4 of $ A $\;
    			 $I3=I3\times I2 $
			}
			$\ M=I3 $\; 
      \end{algorithm}
    \end{minipage}%
    }
     
  \end{center}


\begin{thebibliography}{10}
\expandafter\ifx\csname url\endcsname\relax
  \def\url#1{\texttt{#1}}\fi
\expandafter\ifx\csname urlprefix\endcsname\relax\def\urlprefix{URL }\fi
\expandafter\ifx\csname href\endcsname\relax
  \def\href#1#2{#2} \def\path#1{#1}\fi

\bibitem{hand2001principles}
D.~J. Hand, H.~Mannila, P.~Smyth, Principles of data mining, MIT press, 2001.

\bibitem{bramer2016principles}
M.~Bramer, Principles of data mining, Springer, 2016.

\bibitem{xiao2014data}
F.~Xiao, C.~Fan, Data mining in building automation system for improving
  building operational performance, Energy and buildings 75 (2014) 109--118.

\bibitem{linoff2011data}
G.~S. Linoff, M.~J. Berry, Data mining techniques: for marketing, sales, and
  customer relationship management, John Wiley \& Sons, 2011.

\bibitem{rygielski2002data}
C.~Rygielski, J.-C. Wang, D.~C. Yen, Data mining techniques for customer
  relationship management, Technology in society 24~(4) (2002) 483--502.

\bibitem{du2008recent}
D.~H. Du, Recent advancements and future challenges of storage systems,
  Proceedings of the IEEE 96~(11) (2008) 1875--1886.

\bibitem{wang2014bigdatabench}
L.~Wang, J.~Zhan, C.~Luo, Y.~Zhu, Q.~Yang, Y.~He, W.~Gao, Z.~Jia, Y.~Shi,
  S.~Zhang, et~al., Bigdatabench: A big data benchmark suite from internet
  services, in: High Performance Computer Architecture (HPCA), 2014 IEEE 20th
  International Symposium on, IEEE, 2014, pp. 488--499.

\bibitem{white2012hadoop}
T.~White, Hadoop: The definitive guide, " O'Reilly Media, Inc.", 2012.

\bibitem{karau2015learning}
H.~Karau, A.~Konwinski, P.~Wendell, M.~Zaharia, Learning spark: lightning-fast
  big data analysis, " O'Reilly Media, Inc.", 2015.

\bibitem{carbone2015apache}
P.~Carbone, A.~Katsifodimos, S.~Ewen, V.~Markl, S.~Haridi, K.~Tzoumas, Apache
  flink: Stream and batch processing in a single engine, Bulletin of the IEEE
  Computer Society Technical Committee on Data Engineering 36~(4).

\bibitem{noauthor_big_2014}
\href{https://www.smartdatacollective.com/big-data-20-free-big-data-sources-everyone-should-know/}{Big
  {Data}: 20 {Free} {Big} {Data} {Sources} {Everyone} {Should} {Know}} (Sep.
  2014).
\newline\urlprefix\url{https://www.smartdatacollective.com/big-data-20-free-big-data-sources-everyone-should-know/}

\bibitem{ganti2011mobile}
R.~K. Ganti, F.~Ye, H.~Lei, Mobile crowdsensing: current state and future
  challenges, IEEE Communications Magazine 49~(11).

\bibitem{kim1995masking}
J.~J. Kim, W.~E. Winkler, et~al., Masking microdata files, in: Proceedings of
  the Survey Research Methods Section, American Statistical Association,
  Citeseer, 1995.

\bibitem{rubin1993statistical}
D.~B. Rubin, Statistical disclosure limitation, Journal of official Statistics
  9~(2) (1993) 461--468.

\bibitem{aggarwal2008general}
C.~C. Aggarwal, S.~Y. Philip, A general survey of privacy-preserving data
  mining models and algorithms, in: Privacy-preserving data mining, Springer,
  2008, pp. 11--52.

\bibitem{lu2014toward}
R.~Lu, H.~Zhu, X.~Liu, J.~K. Liu, J.~Shao, Toward efficient and
  privacy-preserving computing in big data era, IEEE Network 28~(4) (2014)
  46--50.

\bibitem{aldeen2015comprehensive}
Y.~A. A.~S. Aldeen, M.~Salleh, M.~A. Razzaque, A comprehensive review on
  privacy preserving data mining, SpringerPlus 4~(1) (2015) 694.

\bibitem{kargupta2003privacy}
H.~Kargupta, S.~Datta, Q.~Wang, K.~Sivakumar, On the privacy preserving
  properties of random data perturbation techniques, in: Data Mining, 2003.
  ICDM 2003. Third IEEE International Conference on, IEEE, 2003, pp. 99--106.

\bibitem{muralidhar1999general}
K.~Muralidhar, R.~Parsa, R.~Sarathy, A general additive data perturbation
  method for database security, management science 45~(10) (1999) 1399--1415.

\bibitem{reiss1980practical}
S.~P. Reiss, Practical data-swapping: The first steps, in: Security and
  Privacy, 1980 IEEE Symposium on, IEEE, 1980, pp. 38--38.

\bibitem{chen2005random}
K.~Chen, L.~Liu, A random rotation perturbation approach to privacy preserving
  data classification.

\bibitem{chen2011geometric}
K.~Chen, L.~Liu, Geometric data perturbation for privacy preserving outsourced
  data mining, Knowledge and Information Systems 29~(3) (2011) 657--695.

\bibitem{aggarwal2004condensation}
C.~C. Aggarwal, P.~S. Yu, A condensation approach to privacy preserving data
  mining, in: EDBT, Vol.~4, Springer, 2004, pp. 183--199.

\bibitem{david_aha_and_fellow_graduate_students_at_uc_irvine_uci_1987}
D.~A.~a. fellow graduate students~at UC~Irvine, A.~Asuncion, D.~Newman,
  \href{https://archive.ics.uci.edu/ml/index.php}{{UCI} {Machine} {Learning}
  {Repository}} (1987).
\newline\urlprefix\url{https://archive.ics.uci.edu/ml/index.php}

\bibitem{witten2016data}
I.~H. Witten, E.~Frank, M.~A. Hall, C.~J. Pal, Data Mining: Practical machine
  learning tools and techniques, Morgan Kaufmann, 2016.

\bibitem{bertino2008survey}
E.~Bertino, D.~Lin, W.~Jiang, A survey of quantification of privacy preserving
  data mining algorithms, in: Privacy-preserving data mining, Springer, 2008,
  pp. 183--205.

\bibitem{verykios2004state}
V.~S. Verykios, E.~Bertino, I.~N. Fovino, L.~P. Provenza, Y.~Saygin,
  Y.~Theodoridis, State-of-the-art in privacy preserving data mining, ACM
  Sigmod Record 33~(1) (2004) 50--57.

\bibitem{domingo2002practical}
J.~Domingo-Ferrer, J.~M. Mateo-Sanz, Practical data-oriented microaggregation
  for statistical disclosure control, IEEE Transactions on Knowledge and data
  Engineering 14~(1) (2002) 189--201.

\bibitem{liu2006random}
K.~Liu, H.~Kargupta, J.~Ryan, Random projection-based multiplicative data
  perturbation for privacy preserving distributed data mining, IEEE
  Transactions on knowledge and Data Engineering 18~(1) (2006) 92--106.

\bibitem{agrawal2000privacy}
R.~Agrawal, R.~Srikant, Privacy-preserving data mining, in: ACM Sigmod Record,
  Vol.~29, ACM, 2000, pp. 439--450.

\bibitem{datta2004random}
S.~Datta, On random additive perturbation for privacy preserving data mining,
  Ph.D. thesis, University of Maryland, Baltimore County (2004).

\bibitem{zhong2012mu}
J.~Zhong, V.~Mirchandani, P.~Bertok, J.~Harland, $\mu$-fractal based data
  perturbation algorithm for privacy protection., in: PACIS, 2012, p. 148.

\bibitem{du2003using}
W.~Du, Z.~Zhan, Using randomized response techniques for privacy-preserving
  data mining, in: Proceedings of the ninth ACM SIGKDD international conference
  on Knowledge discovery and data mining, ACM, 2003, pp. 505--510.

\bibitem{estivill1999data}
V.~Estivill-Castro, L.~Brankovic, Data swapping: Balancing privacy against
  precision in mining for logic rules, in: DaWaK, Vol.~99, Springer, 1999, pp.
  389--398.

\bibitem{defays1998masking}
D.~Defays, M.~Anwar, Masking microdata using micro-aggregation, Journal of
  Official Statistics 14~(4) (1998) 449.

\bibitem{huang2005deriving}
Z.~Huang, W.~Du, B.~Chen, Deriving private information from randomized data,
  in: Proceedings of the 2005 ACM SIGMOD international conference on Management
  of data, ACM, 2005, pp. 37--48.

\bibitem{warner1965randomized}
S.~L. Warner, Randomized response: A survey technique for eliminating evasive
  answer bias, Journal of the American Statistical Association 60~(309) (1965)
  63--69.

\bibitem{liu2007multiplicative}
K.~Liu, Multiplicative data perturbation for privacy preserving data mining,
  Ph.D. thesis, University of Maryland, Baltimore County (2007).

\bibitem{singh2013attack}
K.~Singh, L.~Batten, An attack-resistant hybrid data-privatization method with
  low information loss, in: IFIP International Conference on Trust Management,
  Springer, 2013, pp. 263--271.

\bibitem{weisstein2003chebyshev}
E.~W. Weisstein, Chebyshev polynomial of the first kind.

\bibitem{wold1987principal}
S.~Wold, K.~Esbensen, P.~Geladi, Principal component analysis, Chemometrics and
  intelligent laboratory systems 2~(1-3) (1987) 37--52.

\bibitem{scholz1985maximum}
F.~Scholz, Maximum likelihood estimation, Encyclopedia of statistical sciences.

\bibitem{chen2005privacy}
K.~Chen, L.~Liu, Privacy preserving data classification with rotation
  perturbation, in: Data Mining, Fifth IEEE International Conference on, IEEE,
  2005, pp. 4--pp.

\bibitem{machanavajjhala2006diversity}
A.~Machanavajjhala, J.~Gehrke, D.~Kifer, M.~Venkitasubramaniam, l-diversity:
  Privacy beyond k-anonymity, in: Data Engineering, 2006. ICDE'06. Proceedings
  of the 22nd International Conference on, IEEE, 2006, pp. 24--24.

\bibitem{aggarwal2005k}
C.~C. Aggarwal, On k-anonymity and the curse of dimensionality, in: Proceedings
  of the 31st international conference on Very large data bases, VLDB
  Endowment, 2005, pp. 901--909.

\bibitem{lessmann2015benchmarking}
S.~Lessmann, B.~Baesens, H.-V. Seow, L.~C. Thomas, Benchmarking
  state-of-the-art classification algorithms for credit scoring: An update of
  research, European Journal of Operational Research 247~(1) (2015) 124--136.

\bibitem{scholkopf1999advances}
B.~Sch{\"o}lkopf, C.~J. Burges, A.~J. Smola, Advances in kernel methods:
  support vector learning, MIT press, 1999.

\bibitem{quinlan1993c4}
J.~R. Quinlan, C4. 5: Programming for machine learning, Morgan Kauffmann 38.

\bibitem{howell2016fundamental}
D.~C. Howell, Fundamental statistics for the behavioral sciences, Nelson
  Education, 2016.

\bibitem{okkalioglu2015survey}
B.~D. Okkalioglu, M.~Okkalioglu, M.~Koc, H.~Polat, A survey: deriving private
  information from perturbed data, Artificial Intelligence Review 44~(4) (2015)
  547--569.

\bibitem{gavert2005fastica}
H.~G{\"a}vert, J.~Hurri, J.~S{\"a}rel{\"a}, A.~Hyv{\"a}rinen, The fastica
  package for matlab, Lab Comput Inf Sci Helsinki Univ. Technol.

\end{thebibliography}


\begin{thebibliography}{10}
\expandafter\ifx\csname url\endcsname\relax
  \def\url#1{\texttt{#1}}\fi
\expandafter\ifx\csname urlprefix\endcsname\relax\def\urlprefix{URL }\fi
\expandafter\ifx\csname href\endcsname\relax
  \def\href#1#2{#2} \def\path#1{#1}\fi

\bibitem{tegegne2014enriching}
T.~Tegegne, T.~P.~T. van~der Weide, Enriching queries with user preferences in
  healthcare, Information Processing \& Management 50~(4) (2014) 599--620.

\bibitem{kim2017information}
K.~J. Kim, D.-H. Shin, H.~Yoon, Information tailoring and framing in wearable
  health communication, Information Processing \& Management 53~(2) (2017)
  351--358.

\bibitem{serban2019real}
O.~Șerban, N.~Thapen, B.~Maginnis, C.~Hankin, V.~Foot, Real-time processing of
  social media with sentinel: a syndromic surveillance system incorporating
  deep learning for health classification, Information Processing \& Management
  56~(3) (2019) 1166--1184.

\bibitem{khan2018iot}
M.~A. Khan, K.~Salah, Iot security: Review, blockchain solutions, and open
  challenges, Future Generation Computer Systems 82 (2018) 395--411.

\bibitem{arachchige2020trustworthy}
P.~C.~M. Arachchige, P.~Bertok, I.~Khalil, D.~Liu, S.~Camtepe, M.~Atiquzzaman,
  A trustworthy privacy preserving framework for machine learning in industrial
  iot systems, IEEE Transactions on Industrial Informatics.

\bibitem{arachchige2019local}
P.~C.~M. Arachchige, P.~Bertok, I.~Khalil, D.~Liu, S.~Camtepe, M.~Atiquzzaman,
  Local differential privacy for deep learning, IEEE Internet of Things
  Journal.

\bibitem{chamikara2019efficient}
M.~Chamikara, P.~Bertok, D.~Liu, S.~Camtepe, I.~Khalil, An efficient and
  scalable privacy preserving algorithm for big data and data streams,
  Computers \& Security 87 (2019) 101570.

\bibitem{chamikara2016fuzzy}
M.~A.~P. Chamikara, A.~Galappaththi, R.~D. Yapa, R.~D. Nawarathna, S.~R.
  Kodituwakku, J.~Gunatilake, A.~A. C.~A. Jayathilake, L.~Liyanage, Fuzzy based
  binary feature profiling for modus operandi analysis, PeerJ Computer Science
  2 (2016) e65.

\bibitem{alabdulatif2018real}
A.~Alabdulatif, I.~Khalil, A.~R.~M. Forkan, M.~Atiquzzaman, Real-time secure
  health surveillance for smarter health communities, IEEE Communications
  Magazine 57~(1) (2018) 122--129.

\bibitem{alabdulatif2019secure}
A.~Alabdulatif, I.~Khalil, X.~Yi, M.~Guizani, Secure edge of things for smart
  healthcare surveillance framework, IEEE Access 7 (2019) 31010--31021.

\bibitem{bonawitz2019towards}
K.~Bonawitz, H.~Eichner, W.~Grieskamp, D.~Huba, A.~Ingerman, V.~Ivanov,
  C.~Kiddon, J.~Konecny, S.~Mazzocchi, H.~B. McMahan, et~al., Towards federated
  learning at scale: System design, arXiv preprint arXiv:1902.01046.

\bibitem{bertino2008survey}
E.~Bertino, D.~Lin, W.~Jiang, A survey of quantification of privacy preserving
  data mining algorithms, in: Privacy-preserving data mining, Springer, 2008,
  pp. 183--205.

\bibitem{samarati2001protecting}
P.~Samarati, Protecting respondents identities in microdata release, IEEE
  transactions on Knowledge and Data Engineering 13~(6) (2001) 1010--1027.

\bibitem{chamikara2018efficient}
M.~Chamikara, P.~Bertok, D.~Liu, S.~Camtepe, I.~Khalil, Efficient data
  perturbation for privacy preserving and accurate data stream mining,
  Pervasive and Mobile Computing 48 (2018) 1--19.

\bibitem{lopez2013privacy}
N.~L{\'o}pez, F.~Seb{\'e}, Privacy preserving release of blogosphere data in
  the presence of search engines, Information Processing \& Management 49~(4)
  (2013) 833--851.

\bibitem{bilge2013scalable}
A.~Bilge, H.~Polat, A scalable privacy-preserving recommendation scheme via
  bisecting k-means clustering, Information Processing \& Management 49~(4)
  (2013) 912--927.

\bibitem{li2020voluntary}
K.~Li, L.~Cheng, C.-I. Teng, Voluntary sharing and mandatory provision: Private
  information disclosure on social networking sites, Information Processing \&
  Management 57~(1) (2020) 102128.

\bibitem{zhou2017security}
J.~Zhou, Z.~Cao, X.~Dong, A.~V. Vasilakos, Security and privacy for cloud-based
  iot: Challenges, IEEE Communications Magazine 55~(1) (2017) 26--33.

\bibitem{yargic2019privacy}
A.~Yargic, A.~Bilge, Privacy-preserving multi-criteria collaborative filtering,
  Information Processing \& Management 56~(3) (2019) 994--1009.

\bibitem{yang2019federated}
Q.~Yang, Y.~Liu, T.~Chen, Y.~Tong, Federated machine learning: Concept and
  applications, ACM Transactions on Intelligent Systems and Technology (TIST)
  10~(2) (2019) 12.

\bibitem{thapa2020splitfed}
C.~Thapa, M.~A.~P. Chamikara, S.~Camtepe, Splitfed: When federated learning
  meets split learning, arXiv preprint arXiv:2004.12088.

\bibitem{song2017machine}
C.~Song, T.~Ristenpart, V.~Shmatikov, Machine learning models that remember too
  much, in: Proceedings of the 2017 ACM SIGSAC Conference on Computer and
  Communications Security, ACM, 2017, pp. 587--601.

\bibitem{shokri2017membership}
R.~Shokri, M.~Stronati, C.~Song, V.~Shmatikov, Membership inference attacks
  against machine learning models, in: Security and Privacy (SP), 2017 IEEE
  Symposium on, IEEE, 2017, pp. 3--18.
\newblock \href {http://dx.doi.org/https://doi.org/10.1109/SP.2017.41}
  {\path{doi:https://doi.org/10.1109/SP.2017.41}}.

\bibitem{fredrikson2015model}
M.~Fredrikson, S.~Jha, T.~Ristenpart, Model inversion attacks that exploit
  confidence information and basic countermeasures, in: Proceedings of the 22nd
  ACM SIGSAC Conference on Computer and Communications Security, ACM, 2015, pp.
  1322--1333.

\bibitem{akgun2015privacy}
M.~Akg{\"u}n, A.~O. Bayrak, B.~Ozer, M.~{\c{S}}. Sa{\u{g}}{\i}ro{\u{g}}lu,
  Privacy preserving processing of genomic data: A survey, Journal of
  biomedical informatics 56 (2015) 103--111.

\bibitem{chen2005random}
K.~Chen, L.~Liu, \href{https://corescholar.libraries.wright.edu/knoesis/916/}{A
  random rotation perturbation approach to privacy preserving data
  classification}, The Ohio Center of Excellence in Knowledge-Enabled
  Computing.
\newline\urlprefix\url{https://corescholar.libraries.wright.edu/knoesis/916/}

\bibitem{chen2011geometric}
K.~Chen, L.~Liu, Geometric data perturbation for privacy preserving outsourced
  data mining, Knowledge and Information Systems 29~(3) (2011) 657--695.
\newblock \href {http://dx.doi.org/https://doi.org/10.1007/s10115-010-0362-4}
  {\path{doi:https://doi.org/10.1007/s10115-010-0362-4}}.

\bibitem{okkalioglu2015survey}
B.~D. Okkalioglu, M.~Okkalioglu, M.~Koc, H.~Polat, A survey: deriving private
  information from perturbed data, Artificial Intelligence Review 44~(4) (2015)
  547--569.
\newblock \href {http://dx.doi.org/https://doi.org/10.1007/s10462-015-9439-5}
  {\path{doi:https://doi.org/10.1007/s10462-015-9439-5}}.

\bibitem{chamikara2020efficient}
M.~A.~P. Chamikara, P.~Bert{\'o}k, D.~Liu, S.~Camtepe, I.~Khalil, Efficient
  privacy preservation of big data for accurate data mining, Information
  Sciences 527 (2020) 420--443.

\bibitem{oleshchuk2009internet}
V.~Oleshchuk, Internet of things and privacy preserving technologies, in: 2009
  1st International Conference on Wireless Communication, Vehicular Technology,
  Information Theory and Aerospace \& Electronic Systems Technology, IEEE,
  2009, pp. 336--340.

\bibitem{hasan2016effective}
A.~Hasan, Q.~Jiang, J.~Luo, C.~Li, L.~Chen, An effective value swapping method
  for privacy preserving data publishing, Security and Communication Networks
  9~(16) (2016) 3219--3228.
\newblock \href {http://dx.doi.org/https://doi.org/10.1002/sec.1527}
  {\path{doi:https://doi.org/10.1002/sec.1527}}.

\bibitem{muralidhar1999general}
K.~Muralidhar, R.~Parsa, R.~Sarathy, A general additive data perturbation
  method for database security, management science 45~(10) (1999) 1399--1415.
\newblock \href {http://dx.doi.org/https://doi.org/10.1287/mnsc.45.10.1399}
  {\path{doi:https://doi.org/10.1287/mnsc.45.10.1399}}.

\bibitem{aggarwal2004condensation}
C.~C. Aggarwal, P.~S. Yu, A condensation approach to privacy preserving data
  mining, in: EDBT, Vol.~4, Springer, 2004, pp. 183--199.
\newblock \href
  {http://dx.doi.org/https://doi.org/10.1007/978-3-540-24741-8_12}
  {\path{doi:https://doi.org/10.1007/978-3-540-24741-8_12}}.

\bibitem{fox2015randomized}
J.~A. Fox, \href{https://books.google.com.au/books?isbn=1483381056}{Randomized
  response and related methods: Surveying Sensitive Data}, Vol.~58, SAGE
  Publications, 2015.
\newline\urlprefix\url{https://books.google.com.au/books?isbn=1483381056}

\bibitem{soria2015t}
J.~Soria-Comas, J.~Domingo-Ferrer, D.~S{\'a}nchez, S.~Mart{\'\i}nez,
  t-closeness through microaggregation: Strict privacy with enhanced utility
  preservation, IEEE Transactions on Knowledge and Data Engineering 27~(11)
  (2015) 3098--3110.
\newblock \href {http://dx.doi.org/https://doi.org/10.1109/TKDE.2015.2435777}
  {\path{doi:https://doi.org/10.1109/TKDE.2015.2435777}}.

\bibitem{liu2006random}
K.~Liu, H.~Kargupta, J.~Ryan, Random projection-based multiplicative data
  perturbation for privacy preserving distributed data mining, IEEE
  Transactions on knowledge and Data Engineering 18~(1) (2006) 92--106.
\newblock \href {http://dx.doi.org/https://doi.org/10.1109/TKDE.2006.14}
  {\path{doi:https://doi.org/10.1109/TKDE.2006.14}}.

\bibitem{aldeen2015comprehensive}
Y.~A. A.~S. Aldeen, M.~Salleh, M.~A. Razzaque, A comprehensive review on
  privacy preserving data mining, SpringerPlus 4~(1) (2015) 694.
\newblock \href {http://dx.doi.org/https://doi.org/10.1186/s40064-015-1481-x}
  {\path{doi:https://doi.org/10.1186/s40064-015-1481-x}}.

\bibitem{machanavajjhala2015designing}
A.~Machanavajjhala, D.~Kifer, Designing statistical privacy for your data,
  Communications of the ACM 58~(3) (2015) 58--67.
\newblock \href {http://dx.doi.org/https://doi.org/10.1145/2660766}
  {\path{doi:https://doi.org/10.1145/2660766}}.

\bibitem{niu2014achieving}
B.~Niu, Q.~Li, X.~Zhu, G.~Cao, H.~Li, Achieving k-anonymity in privacy-aware
  location-based services, in: INFOCOM, 2014 Proceedings IEEE, IEEE, 2014, pp.
  754--762.
\newblock \href
  {http://dx.doi.org/https://doi.org/10.1109/INFOCOM.2014.6848002}
  {\path{doi:https://doi.org/10.1109/INFOCOM.2014.6848002}}.

\bibitem{navarro2012user}
G.~Navarro-Arribas, V.~Torra, A.~Erola, J.~Castell{\`a}-Roca, User k-anonymity
  for privacy preserving data mining of query logs, Information Processing \&
  Management 48~(3) (2012) 476--487.

\bibitem{machanavajjhala2006diversity}
A.~Machanavajjhala, J.~Gehrke, D.~Kifer, M.~Venkitasubramaniam, l-diversity:
  Privacy beyond k-anonymity, in: Data Engineering, 2006. ICDE'06. Proceedings
  of the 22nd International Conference on, IEEE, 2006, pp. 24--24.
\newblock \href {http://dx.doi.org/https://doi.org/10.1109/ICDE.2006.1}
  {\path{doi:https://doi.org/10.1109/ICDE.2006.1}}.

\bibitem{li2007t}
N.~Li, T.~Li, S.~Venkatasubramanian, t-closeness: Privacy beyond k-anonymity
  and l-diversity, in: Data Engineering, 2007. ICDE 2007. IEEE 23rd
  International Conference on, IEEE, 2007, pp. 106--115.
\newblock \href {http://dx.doi.org/https://doi.org/10.1109/ICDE.2007.367856}
  {\path{doi:https://doi.org/10.1109/ICDE.2007.367856}}.

\bibitem{wong2006alpha}
R.~C.-W. Wong, J.~Li, A.~W.-C. Fu, K.~Wang, ($\alpha$, k)-anonymity: an
  enhanced k-anonymity model for privacy preserving data publishing, in:
  Proceedings of the 12th ACM SIGKDD international conference on Knowledge
  discovery and data mining, ACM, 2006, pp. 754--759.
\newblock \href {http://dx.doi.org/https://doi.org/10.1145/1150402.1150499}
  {\path{doi:https://doi.org/10.1145/1150402.1150499}}.

\bibitem{carpineto2015ktheta}
C.~Carpineto, G.~Romano, K$\theta$-affinity privacy: Releasing infrequent query
  refinements safely, Information Processing \& Management 51~(2) (2015)
  74--88.

\bibitem{ganta2008composition}
S.~R. Ganta, S.~P. Kasiviswanathan, A.~Smith, Composition attacks and auxiliary
  information in data privacy, in: Proceedings of the 14th ACM SIGKDD
  international conference on Knowledge discovery and data mining, ACM, 2008,
  pp. 265--273.
\newblock \href {http://dx.doi.org/https://doi.org/10.1145/1401890.1401926}
  {\path{doi:https://doi.org/10.1145/1401890.1401926}}.

\bibitem{zhang2007information}
L.~Zhang, S.~Jajodia, A.~Brodsky, Information disclosure under realistic
  assumptions: Privacy versus optimality, in: Proceedings of the 14th ACM
  conference on Computer and communications security, ACM, 2007, pp. 573--583.
\newblock \href {http://dx.doi.org/https://doi.org/10.1145/1315245.1315316}
  {\path{doi:https://doi.org/10.1145/1315245.1315316}}.

\bibitem{wong2011can}
R.~C.-W. Wong, A.~W.-C. Fu, K.~Wang, P.~S. Yu, J.~Pei, Can the utility of
  anonymized data be used for privacy breaches?, ACM Transactions on Knowledge
  Discovery from Data (TKDD) 5~(3) (2011) 16.
\newblock \href {http://dx.doi.org/https://doi.org/10.1145/1993077.1993080}
  {\path{doi:https://doi.org/10.1145/1993077.1993080}}.

\bibitem{aggarwal2008privacy}
C.~C. Aggarwal, Privacy and the dimensionality curse, Privacy-Preserving Data
  Mining (2008) 433--460.\href
  {http://dx.doi.org/https://doi.org/10.1007/978-0-387-70992-5_18}
  {\path{doi:https://doi.org/10.1007/978-0-387-70992-5_18}}.

\bibitem{bettini2015privacy}
C.~Bettini, D.~Riboni, Privacy protection in pervasive systems: State of the
  art and technical challenges, Pervasive and Mobile Computing 17 (2015)
  159--174.
\newblock \href {http://dx.doi.org/https://doi.org/10.1016/j.pmcj.2014.09.010}
  {\path{doi:https://doi.org/10.1016/j.pmcj.2014.09.010}}.

\bibitem{hardy2017private}
S.~Hardy, W.~Henecka, H.~Ivey-Law, R.~Nock, G.~Patrini, G.~Smith, B.~Thorne,
  Private federated learning on vertically partitioned data via entity
  resolution and additively homomorphic encryption, arXiv preprint
  arXiv:1711.10677.

\bibitem{maruskin2012essential}
J.~Maruskin, \href{https://books.google.com.au/books?id=aOF3-hx3u1kC}{Essential
  Linear Algebra}, Solar Crest Publishing, LLC, 2012.
\newline\urlprefix\url{https://books.google.com.au/books?id=aOF3-hx3u1kC}

\bibitem{jones2012computer}
H.~Jones, \href{https://books.google.com.au/books?id=f7gPBwAAQBAJ}{Computer
  Graphics through Key Mathematics}, Springer London : Imprint: Springer, 2012.
\newline\urlprefix\url{https://books.google.com.au/books?id=f7gPBwAAQBAJ}

\bibitem{kabir2015novel}
W.~Kabir, M.~O. Ahmad, M.~Swamy, A novel normalization technique for multimodal
  biometric systems, in: Circuits and Systems (MWSCAS), 2015 IEEE 58th
  International Midwest Symposium on, IEEE, 2015, pp. 1--4.
\newblock \href {http://dx.doi.org/https://doi.org/10.1109/MWSCAS.2015.7282214}
  {\path{doi:https://doi.org/10.1109/MWSCAS.2015.7282214}}.

\bibitem{bennett2009numerically}
J.~Bennett, R.~Grout, P.~P{\'e}bay, D.~Roe, D.~Thompson, Numerically stable,
  single-pass, parallel statistics algorithms, in: Cluster Computing and
  Workshops, 2009. CLUSTER'09. IEEE International Conference on, IEEE, 2009,
  pp. 1--8.

\bibitem{witten2016data}
I.~H. Witten, E.~Frank, M.~A. Hall, C.~J. Pal,
  \href{https://books.google.com.au/books?isbn=0128043571}{Data Mining:
  Practical machine learning tools and techniques}, Morgan Kaufmann, 2016.
\newline\urlprefix\url{https://books.google.com.au/books?isbn=0128043571}

\bibitem{leon2016controlling}
A.~S. Leon, C.~Goodell, Controlling hec-ras using matlab, Environmental
  modelling \& software 84 (2016) 339--348.

\bibitem{wang2019adaptive}
S.~Wang, T.~Tuor, T.~Salonidis, K.~K. Leung, C.~Makaya, T.~He, K.~Chan,
  Adaptive federated learning in resource constrained edge computing systems,
  IEEE Journal on Selected Areas in Communications 37~(6) (2019) 1205--1221.

\bibitem{dinh2019federated}
C.~Dinh, N.~H. Tran, M.~N. Nguyen, C.~S. Hong, W.~Bao, A.~Zomaya, V.~Gramoli,
  Federated learning over wireless networks: Convergence analysis and resource
  allocation, arXiv preprint arXiv:1910.13067.

\bibitem{manogaran2017big}
G.~Manogaran, C.~Thota, D.~Lopez, V.~Vijayakumar, K.~M. Abbas, R.~Sundarsekar,
  Big data knowledge system in healthcare, in: Internet of things and big data
  technologies for next generation healthcare, Springer, 2017, pp. 133--157.
\newblock \href {http://dx.doi.org/https://doi.org/10.1007/978-3-319-49736-5_7}
  {\path{doi:https://doi.org/10.1007/978-3-319-49736-5_7}}.

\bibitem{paeth2014graphics}
A.~W. Paeth, \href{https://books.google.com.au/books?isbn=1483296695}{Graphics
  Gems V (Macintosh Version)}, Academic Press, 2014.
\newline\urlprefix\url{https://books.google.com.au/books?isbn=1483296695}

\end{thebibliography}
\end{document}